\documentclass[12pt,preprint]{aastex}
\usepackage{emulateapj5}
\usepackage{graphicx}
\usepackage{natbib}

\slugcomment{ApJ accepted}

\begin{document}

\title{Non-Thermal X-ray Properties of Rotation Powered Pulsars and Their Wind Nebulae}

\author{Xiang-Hua Li\altaffilmark{1,2}, Fang-Jun Lu\altaffilmark{1} and Zhuo Li\altaffilmark{3}}

\altaffiltext{1}{Laboratory of Particle Astrophysics, Institute of High Energy Physics,
CAS, Beijing 100049, China; lixh@ihep.ac.cn; lufj@ihep.ac.cn}
\altaffiltext{2}{Max-Planck-Institut f\"{u}r Radioastronomie,
                 Auf dem H\"ugel 69, 53121 Bonn, Germany}
\altaffiltext{3}{Physics Faculty, Weizmann Institute of Science, Rehovot 76100, Israel;
lizhuo@weizmann.ac.il}
\date{Received / Accepted}

\begin{abstract}
We present a statistical study of the non-thermal X-ray emission of 27 young rotation
powered pulsars (RPPs) and 24 pulsar wind nebulae (PWNe) by using the {\sl Chandra} and
the {\sl XMM-Newton} observations, which with the high spatial resolutions enable us to
spatially resolve pulsars from their surrounding PWNe. We obtain the X-ray luminosities
and spectra separately for RPPs and PWNe, and then investigate their distribution and
relation to each other as well as the relation with the pulsar rotational parameters. In
the pair-correlation analysis we find that: (1) the X-ray (2-10 keV) luminosities of both
pulsar and PWN ($L_{\rm X,psr}$ and $L_{\rm X,pwn}$) display a strong correlation with
pulsar spin down power  $\dot{E}$ and characteristic age $\tau$, and the scalings
resulting from a simple linear fit to the data are $L_{\rm
X,psr}\propto\dot{E}^{0.92\pm0.04}$ and $L_{\rm X,pwn}\propto\dot{E}^{1.45\pm0.08}$ (68\%
confidence level), respectively, however, both the fits are not statistically acceptable;
(2) $L_{\rm X,psr}$ also shows a possible weak correlation with pulsar period $P$ and
period derivative $\dot{P}$, whereas $L_{\rm X,pwn}$ manifests a similar weak correlation
with $\dot{P}$ only; (3) The PWN photon index $\Gamma_{\rm pwn}$ is positively correlated
with $L_{\rm X,pwn}$ and $L_{\rm X,pwn}/\dot{E}$. We also found that the PWN X-ray
luminosity is typically 1 to 10 times larger than that from the underlying pulsar, and
the PWN photon indices span a range of $1.5\lesssim\Gamma_{\rm pwn}\lesssim2$. The
statistic study of PWN spectral properties supports the particle wind model in which the
X-ray emitting electrons are accelerated by the termination shock of the wind.
\end{abstract}


\section{Introduction}
The rotation-powered pulsars (RPPs) are known as rapidly spinning and strongly magnetized
neutron stars that are radiating at the expense of their rotational energy. The X-ray
emission of RPPs may contain both thermal and non-thermal components. The thermal emission
might be further divided into non-pulsed and pulsed components. The non-pulsed component,
originates from the cooling of the neutron star,
is from the whole pulsar surface with a characteristic temperature of about 0.1 keV,
 while the pulsed component comes from hot spots
($\sim$ 1.0 keV) on the pulsar surface, which are heated by the bombardment of
relativistic particles streaming back to the surface from the pulsar magnetosphere (Kundt
\& Schaaf 1993, Zavlin et al. 1995, Gil \& Krawczyk 1996). The non-thermal pulsar emission
is from the pulsar magnetosphere, and it might also contain pulsed (e.g., Cheng \& Zhang
1999; Zhang \& Harding 2000) and non-pulsed components (e.g., Tennant et al. 2001; Becker
et al. 2004). In some cases, a pulsar wind nebula (PWN) is found to surround a RPP.  The
X-ray emission of the PWN is non-pulsed and often dominates the non-thermal emission of
the system.

A lot of efforts have been devoted to the statistical studies of pulsar X-ray emission
properties, with particular emphasis on the efficiency of conversion of the pulsar spin
down power $\dot{E}$ into X-ray luminosity. By using the {\sl Einstein} data, Seward \&
Wang (1988) found that $L_{\rm X}\propto \dot{E}^{1.39}$, where $L_{\rm X}$ is the
0.2-4.0 keV X-ray luminosity of the pulsar (plus PWN).  Becker \& Tr{\"u}mper (1997)
obtained $L_{\rm X}\simeq 10^{-3}\dot{E}$ using a sample of 27 pulsars observed by {\sl
ROSAT}, where $L_{\rm X}$ is the total X-ray (0.1-2.4 keV) luminosity of the pulsar plus
PWN. However, in these two works, the thermal emission may contribute significantly to
the total X-ray luminosity given the adopted soft X-ray band. Saito (1998) analyzed 16
RPPs observed by {\sl ASCA} (2-10 keV) and found $L_{\rm X}\propto \dot{E}^{3/2}$.
Possenti et al. (2002) reported $L_{\rm X}\propto \dot{E}^{1.34}$ using 39 pulsars
observed by  {\sl ASCA, RXTE, BeppoSAX, Chandra} and {\sl XMM-Newton}. The X-ray
luminosities in Saito (1998) and Possenti et al. (2002) also include the total emission
due to the pulsars plus PWNe, given the limited spatial resolutions of {\sl ASCA}, {\sl
RXTE}, and {\sl BeppoSAX}. Cheng et al. (2004) divided the total X-ray emission into
pulsed and non-pulsed components, and found that the X-ray luminosity of the pulsed one
follows $L_{\rm X,pul}\propto\dot{E}^{1.2}$, which agrees with the model prediction
$L_{\rm X}\propto\dot{E}^{1.15}$ by Cheng \& Zhang (1999). For non-pulsed emission, they
got $L_{\rm X,npul}\propto\dot{E}^{1.4}$, where they supposed that the non-pulsed
component comes mainly from PWNe and the contribution of non-pulsed component from
pulsars is negligible. It is worth noting that the scatter in the relation is large,
as pointed out by Possenti et al. (2002), who performed study including the estimates of the observational
errors and showing that the linear fit is statistically unacceptable.

All these previous works suffer from the low spatial resolution of the detectors, making
it difficult to resolve the emission of the pulsars from that of their surrounding PWNe.
It is the purpose of our current work to resolve and to analyze the pulsar and the PWN emission
separately. Thanks to the high spatial resolution observations performed with {\sl
Chandra} and {\sl XMM-Newton}, we have been able to satisfactorily investigate 27 pulsars
and 24 PWNe, for which we  have determined the non-thermal X-ray fluxes and spectra in
the 2-10 keV band. Then we have carried out separated statistic studies of RPPs and PWNe,
and tested the consistence of their emission properties with current models. The
organization of this paper is as following: the sample and the data processing are presented
in section 2; the statistical analyses of the X-ray spectral properties of RPPs and PWNe are
given in section 3; we discuss the physical implications of our results in section 4 and
summarize our work in section 5.

\section{Sample and Data processing}
We collect pulsar and PWN samples from the observations by {\sl Chandra} and {\sl
XMM-Newton}, which both have high spatial resolutions, i.e., $\sim1\arcsec$ and
$\sim6\arcsec$, respectively. We take the {\sl Chandra} data directly from the
literatures, and if there are no published results, we analyzed the data in this paper.
The {\sl XMM-Newton} data are adapted only if there are no relevant data from {\sl
Chandra} for the same source. All the {\sl XMM-Newton} results are taken from
literatures. Totally we obtain the X-ray spectra of 27 RPPs and 24 PWNe. In our samples,
millisecond pulsars (MSPs) are not included. It is generally believed that MSPs have ever
undergone an accretion-driven spin-up phase and they are usually old and regarded as a
significantly different class. Similar study on the MSPs is also limited by the rare data
available. Therefore we do not analyze MSPs here, although we discuss them when compare
our analysis with the previous work including MSPs.

In our samples, there are 15, out of 27, spectra of pulsars obtained from the
archived {\sl Chandra} data. We select only the pulsars detected by the Advanced CCD
Imaging Spectrometer (ACIS) in the Timed Exposure (TE) Mode, in which a pulsar is able
to be resolved spatially from its surrounding PWN. We calibrate the data with CIAO
(ver 3.4) and CALDB (ver 3.3.0). We first reprocess the Level 1 data
 for the correction of the charge transfer inefficiency (CTI) effects, then
clean the background and remove the afterglow. Time intervals
with anomalous background rates associated with particle flare events are further
rejected in the Level 2 data. And then the pulsar positions are obtained by the {\it
celldetect} tool in CIAO.  Finally, the spectra are extracted from the Level 2 data and then fit
with {\sl XSPEC}. We use both the power-law (PL) and the power-law+blackbody (PL+BB)
models to fit the pulsar spectra. If the resulted spectral indices are consistent
within errors in both models, then the results from the PL model are used, otherwise
those from the PL+BB model are used. In our spectral analysis, we show errors at the
90\% confidence level.

Pileup occurs when more than one photon are collected in one pixel within a CCD
readout frame, since those photons can only be recorded as a single photon event whose
energy is the sum of the collected photons.  Therefore pileup may affect the
results of spectral analysis. According to section 6.14.2 in the {\sl Chandra} Proposers'
Observatory Guide v.7\footnote{http://cxc.harvard.edu/proposer/POG/}, the effect of
pileup can be omitted if the pileup fraction is $\le$10\%. However, pileup does affect
the spectral analysis even if the pileup fraction is $\le$10\%. For example, since the
pileup fraction of PSR J1930+1852 is estimated to be only 6\%, its spectral index is
reported to be $1.09^{+0.08}_{-0.09}$ without pileup correction (Lu et al. 2002),
whileas the spectral index is $1.35^{+0.06}_{-0.10}$ after pileup correction (Camilo
et al. 2002). In our spectral analysis, we first estimate the pileup fraction using
PIMMS\footnote{http://cxc.harvard.edu/toolkit/pimms.jsp}, and then add a pileup model
in the spectral fitting of the pulsars if the pileup fraction is higher than 3\%.
Totally, there are 8 pulsars in which the pileup model is included in the spectral
fitting, i.e., PSRs J0205+6449, J0537$-$6910, B0540$-$69 (and its PWN), B0833$-$45
(Vela), J1747$-$2958, J1846$-$0258, J1930+1852 and B1951+32.

The absorption column density ($N_{\rm H}$) is obtained in several ways. (1) For 6
pulsars (PSRs J0205+6449, J0537$-$6910, J1747$-$2958, J1846-0258, J1930+1852 and
B1951+32) with bright PWNe, $N_{\rm H}$ are obtained from the spectral fitting of
their PWNe, and then fixed when fitting the spectra of the pulsars. (2)
PSRs B0540$-$69 and J1124$-$5916 are embedded in SNRs 0540$-$69.3 and G292.0+1.8,
respectively, and their PWN spectra below 2.5 keV are strongly affected by the SNRs.
At the same time, constraining $N_{\rm H}$ with emission above 2.5 keV is difficult
because of the small absorption in this high energy range. Therefore their $N_{\rm H}$
are obtained by fitting the pulsar spectra that the contamination of the SNR emission
is negligible, and then $N_{\rm H}$ are fixed in the spectral fitting of their PWNe.
(3) The PWNe associated with PSRs B0355+54, J1617$-$5055, B1823$-$13, B1929+10 and
J2229+6114 are not bright, and $N_{\rm H}$ is determined by jointly fitting the
spectra of both the pulsar and its PWN. (4) PSRs J0633+1746 and B0833$-$45 have been
studied extensively, and their $N_{\rm H}$ values used in our spectral fitting are
taken from Caraveo et al. (2004) and Pavlov et al. (2001).

The luminosity uncertainty is crucial in our analysis of correlations, and
should be considered carefully. Since the X-ray luminosity is given by $L_{\rm X}=4\pi
d^2f_{\rm X}$, where $d$ is the pulsar distance and $f_{\rm X}$ is the 2-10 keV X-ray
flux, the $L_{\rm X}$ uncertainty should be derived from the uncertainties of both
$f_{\rm X}$ and $d$. The uncertainty of $f_{\rm X}$ is derived from those of the
normalization and the photon index in the spectral fitting. For the fluxes taken from
literatures, their uncertainties are extrapolated from the published ones by the
ratios of the fluxes in 2-10 keV to those in the corresponding published energy ranges.

The distances are usually not well constrained, thus the distance uncertainty
may dominate the luminosity uncertainty. There are several cases in our
samples: (1) the distances of 7 pulsars are derived from the radio dispersion
measures, and their errors are conservatively taken to be 40\%, as estimated by Cordes
\& Lazio (2001); (2) the distances of 14 pulsars are obtained via
their associated SNRs, and some of them  are shown with published distance errors in
literatures, while for the others without published errors a conservative error of
50\% is taken; (3) PSRs J0537$-$6910 and B0540$-$69 are both located in the Large
Magellanic Cloud (LMC), whose distance is taken as 50 kpc, and an
error of 10 kpc for a conservative estimation is adapted (Bradley 2007).

We summarize the properties of all the 27 RPPs and 24 PWNe in Tables 1 and
2, respectively. In Table 2 only 22 PWNe are associated with the RPPs listed in
Table 1. The other two pulsars are excluded from Table 1: (1) Camilo et al. (2004)
suggested that PSR J1016$-$5857 is too faint to be resolved from its background PWN in
the {\it Chandra} Observation; (2) Hessels et al. (2004) found that the spectrum of
PSR J2021+3651 can be fit by a BB model and is thus dominated by thermal components.
On the other hand, 5 out of the 27 RPP samples in Table 1 are not listed in Table 2
for PWNe, because the following 5 pulsars have no PWN reported: B0628$-$28,
B0656+14, B0823+26, B0950+08 and B1055$-$52 ({\" O}gelman \& Tepedelenlio{\v g}lu
2004, Becker et al. 2004, De Luca et al. 2005).

\section{Analyses and Results}
\begin{figure}[!htbp]
\includegraphics[scale=.6]{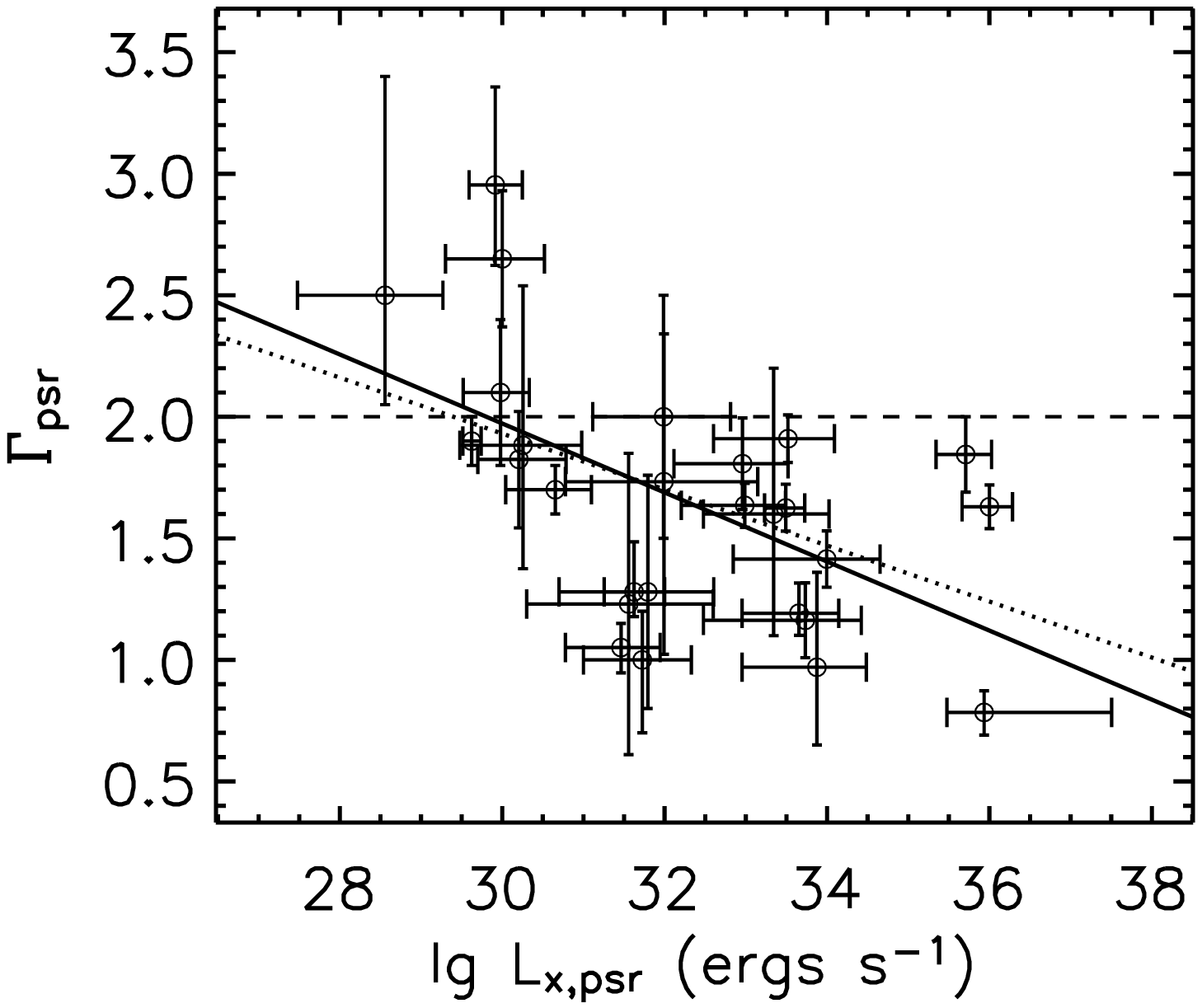}
\includegraphics[scale=.6]{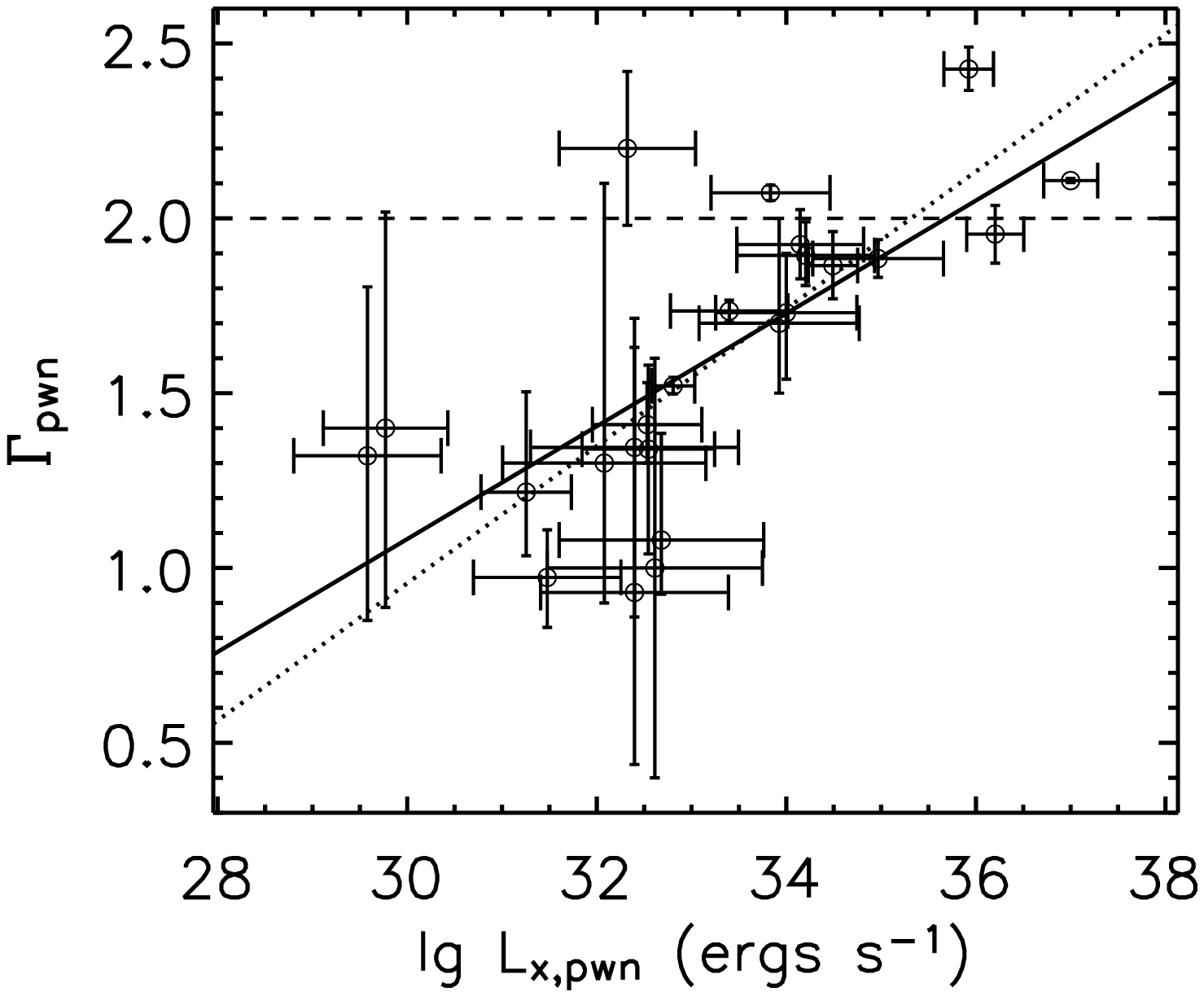}
\caption{ The relations between the non-thermal X-ray luminosity
in 2-10 keV and the photon index for RPPs (left panel) and PWNe (right panel). The
solid lines are the best LSM fit without observational errors taken into account, while the
dotted line with observational errors.
The dashed lines mark the cases of $\Gamma$=2 for comparisons.} \label{fig:Lxvs.Gamma}
\end{figure}

The pulsar photon indices are distributed in a range of $1\la\Gamma_{\rm psr}\la3$ (as
shown in Fig \ref{fig:Lxvs.Gamma}). It should be noted that a significant fraction, about
$\sim15\%$ (4 out of 27), of the sources have soft spectra of $\Gamma_{\rm psr}>2$, which
may raise problems for current models as discussed later in \S 4. The photon indices of
the PWNe span a narrower range (Fig \ref{fig:Lxvs.Gamma}). As discussed below (\S 4),
this is consistent with the pulsar wind model.

We investigate below the correlations between the X-ray emission properties of RPPs
and PWNe, and between the emission properties and the pulsar rotational parameters.
The rotation parameters include the period $P$, the period derivatives $\dot{P}$, and
some derived parameters, e.g., the magnetic field
$B=3.3\times10^{19}(P\dot{P})^{1/2}$G, the characteristic age $\tau=P/2\dot{P}$, and
the spin down power $\dot{E}=4\pi^2I\dot{P}/P^3$, where a typical moment of
$I=10^{45}\rm g\;cm^{2}$ is assumed. We have taken the values of $P$ and $\dot{P}$
from the pulsar catalog by Manchester et al. (2005)\footnote{See
http://www.atnf.csiro.au/research/pulsar/psrcat/}.

In order to evaluate the significance level of the correlations of the two parameters
concerned, we calculate the widely used Pearson correlation coefficient ($r$), the
Spearman rank correlation coefficient ($r_{\rm s}$), and the two-sided significance level
($p_{\rm s}$) of the Spearman rank test. The results are listed in Table 3.

In addition to the correlation tests, we also perform a linear fit using the least
square method (LSM) to the relevant relations of the parameter pairs. Since the fitting
results are usually dominated by a few data points with much smaller observational errors
than the others, we also perform a linear fit without the observational errors for comparison.
The fitting results are all listed in Table 3, and shown in the relevant figures. In the
following we present the results in details.

\subsection{Correlations between the RPP emission properties and the pulsar
rotational parameters}

\begin{figure}
\includegraphics[scale=.6]{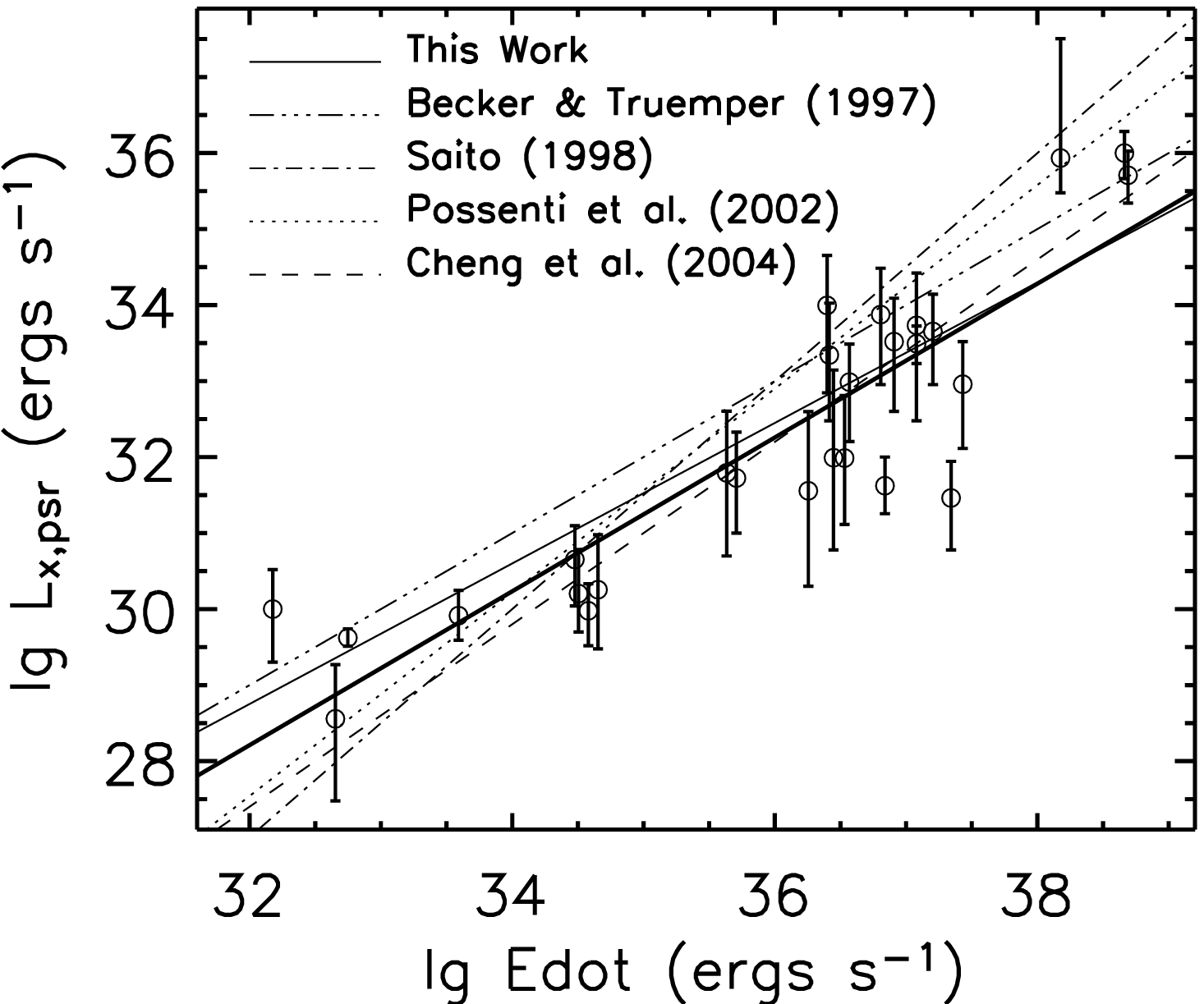}
\includegraphics[scale=.6]{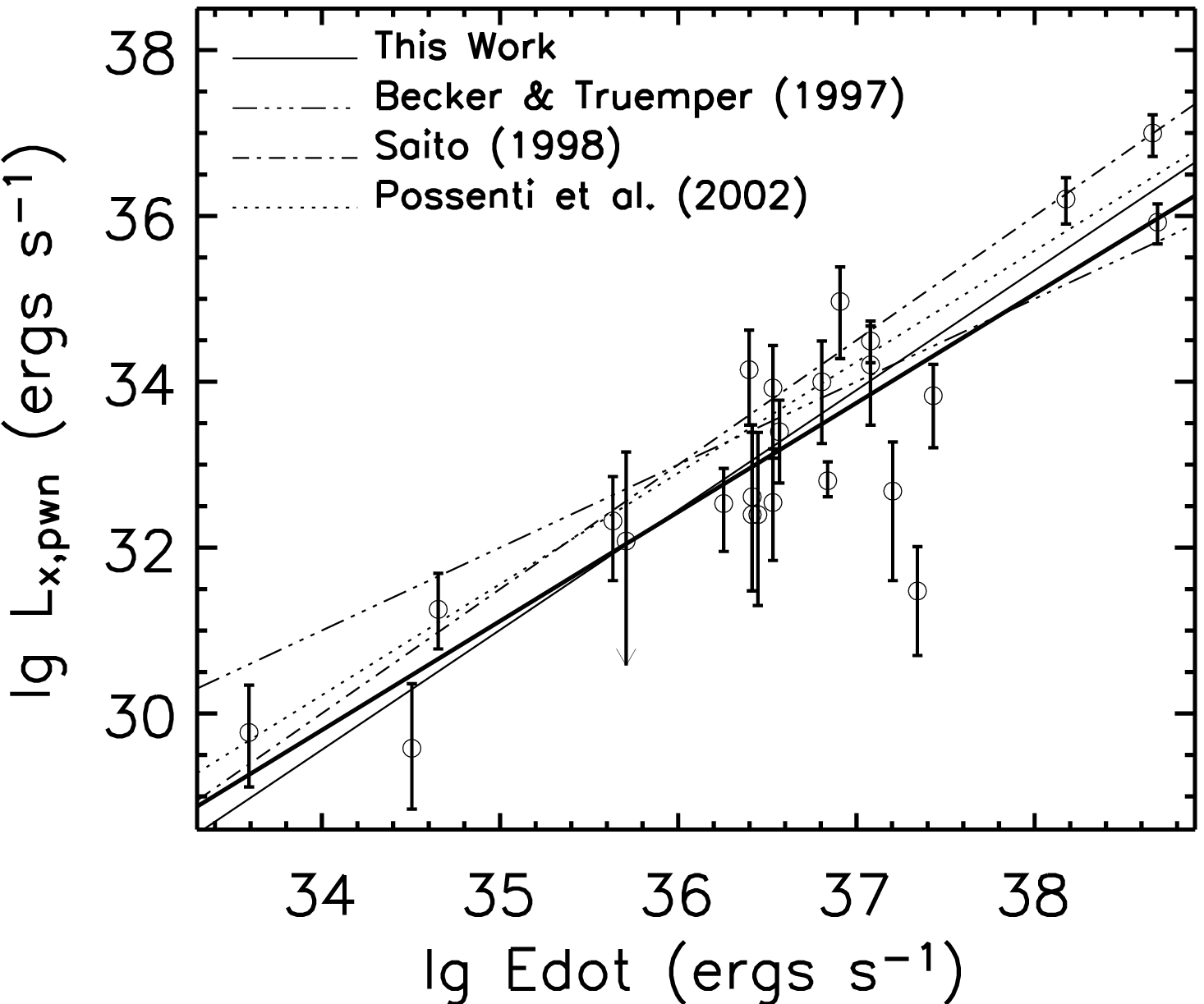}
\caption{ Left: The
non-thermal X-ray luminosities ($L_{\rm X,psr}$) in 2-10 keV from the {\sl Chandra} and {\sl
XMM-Newton} observations vs. spin-down power $\dot{E}$ of 27 pulsars. Right:
The non-thermal X-ray luminosity ($L_{\rm X,pwn}$) in 2-10 keV from the {\sl Chandra} and
{\sl XMM-Newton} observations vs. spin-down power $\dot{E}$ of 24 PWNe.  The bold
solid line is the best fit to the data by LSM without observational errors included, while
the thin solid line is the fit with observational errors.  For
comparison the resulted relations in previous works are also marked: the dash-dot-dot-dot
line corresponds to results of Becker \& Tr{\"u}mper (1997); the dash-dot line to Saito
(1998); the short dash line to Possenti et al. (2002); and long dash line to Cheng et al.
(2004).} \label{fig_le}
\end{figure}

\begin{figure}
\includegraphics[scale=1.]{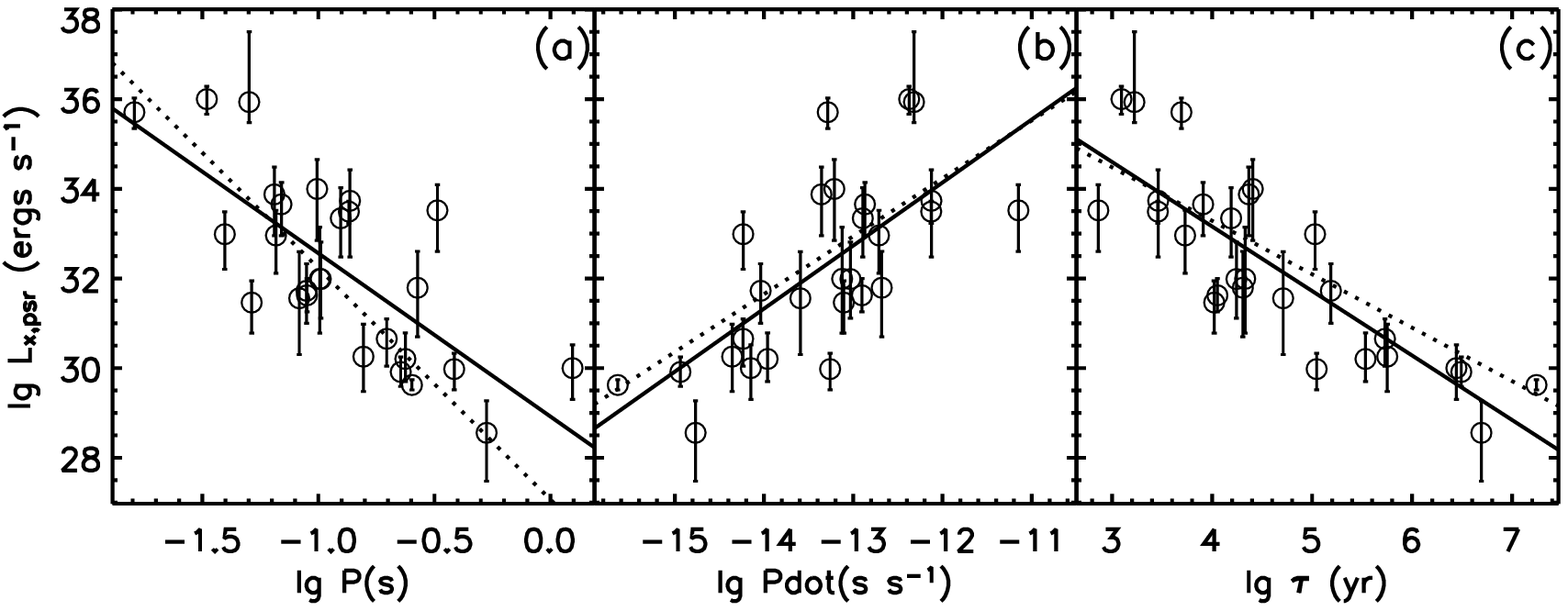}
\caption{Relations of $L_{\rm X,psr}$ vs. $P$, $\dot{P}$ and $\tau$. The solid lines are
the best LSM fit without observational errors considered while the dotted lines are with
errors. The fitting results are also listed in Table 3.} \label{fig_Lx_psr}
\end{figure}

We study the RPP emission first. Strong correlations appear between the X-ray
luminosities of pulsars ($L_{\rm X, psr}$) and the pulsar rotational parameters (see
Table 3 and Figs \ref{fig_le} and \ref{fig_Lx_psr}).
First, $L_{\rm X,psr}$ is negatively correlated with $\tau$ and positively correlated
with $\dot{E}$, which are supported by the Spearman tests: $r_{\rm s} = -0.81$ and
$p_{\rm s}<0.0001$ between $L_{\rm X,psr}$ and $\tau$; and $r_{\rm s}=0.82$ and
$p_{\rm s}<0.0001$ between $L_{\rm X,psr}$ and $\dot{E}$.
We also note that there are some hints of the correlation hold for  $L_{\rm X,psr}$ vs
$P$ and $\dot{P}$ separately, with the relevant Spearman rank correlation coefficients of
$-0.66$ and $0.69$ respectively and both significance levels  $<0.001$.
The correlations between $L_{\rm X,psr}$ vs $P$ and $\dot{P}$  will disappear when the
sample includes both the MSPs and the normal RPPs, just as shown by Possenti et al.
(2002).

Despite the Pearson and Spearman correlation coefficient support the existence of a
correlation between $L_{\rm X,psr}$ and $\dot{E}$ (or $\tau$), a simple linear fit to the
logarithm of the data points with the observational errors included does not produce a
statistically acceptable model. In fact, it results (here $L_{\rm X,psr}$ and $\dot{E}$
are in units of erg s$^{-1}$ and $\tau$ in years; see also Figs 2 and 3).
\[
L_{\rm X,psr}=10^{-0.8\pm1.3}\dot{E}^{0.92\pm0.04}(\chi^2 = 2.6) \]
\[ L_{\rm X,psr} = 10^{38.1\pm0.3}\tau^{-1.19\pm0.05}(\chi^2 = 4.9)
\]
Here and elsewhere in this paper, the uncertainties on the linear fits are reported at
68\% confidence level. Previous authors (notably Possenti et al. 2002) also noticed that
a large scatter in the plot prevents to obtain an acceptable fit of the data with a
simple power law dependence of $L_{\rm X,psr}$ on $\dot{E}$. Hence this relation must
only be seen as an empirical average trend and not suitable for predicting the luminosity
of any specific source.

We have also explored a linear fit which does not account for the uncertainties on the values of $L_{\rm X,psr}$. It turns out
\[
L^{*}_{\rm X,psr}=10^{-4.2\pm3.7}\dot{E}^{1.0\pm0.1} \]
\[ L^{*}_{\rm X,psr} = 10^{38.9\pm0.9}\tau^{-1.4\pm0.2}
\]
(see Figs \ref{fig_le} and~\ref{fig_Lx_psr}). A comparison between the current $L_{\rm
X,psr}$ versus $\dot{E}$ relation and the previous studies is shown in Fig \ref{fig_le}.
It can be seen that the relation we obtain above is close to the one between the pulsed
X-ray emission and the spin down power in Cheng et al. (2004), which indicates that most
of the non-thermal X-ray emission from a pulsar is pulsed.

As already done by Possenti et al. (2002) using a sample including also the MSPs (but not
disentangling PWN from RPP emission), we also try to fit $L_{\rm X}$ with
$aP^{b}\dot{P}^c$. This gives the relation
\[
L_{\rm X,psr}=(40\pm1)P^{-3.4\pm0.3}\dot{P}^{0.77\pm0.07}(\chi^2=2.5).
\]
The nominal result of this (still statistically unacceptable) fit would suggest a
preferred dependence of $L_{\rm X,psr}$ on $\dot{E}/P$: however we note that, accounting
for the uncertainties on the parameters, the simpler dependence on $\dot{E}$ (recovered
in the work of Possenti et al. 2002) is also viable.

\begin{figure}
\includegraphics[scale=.7]{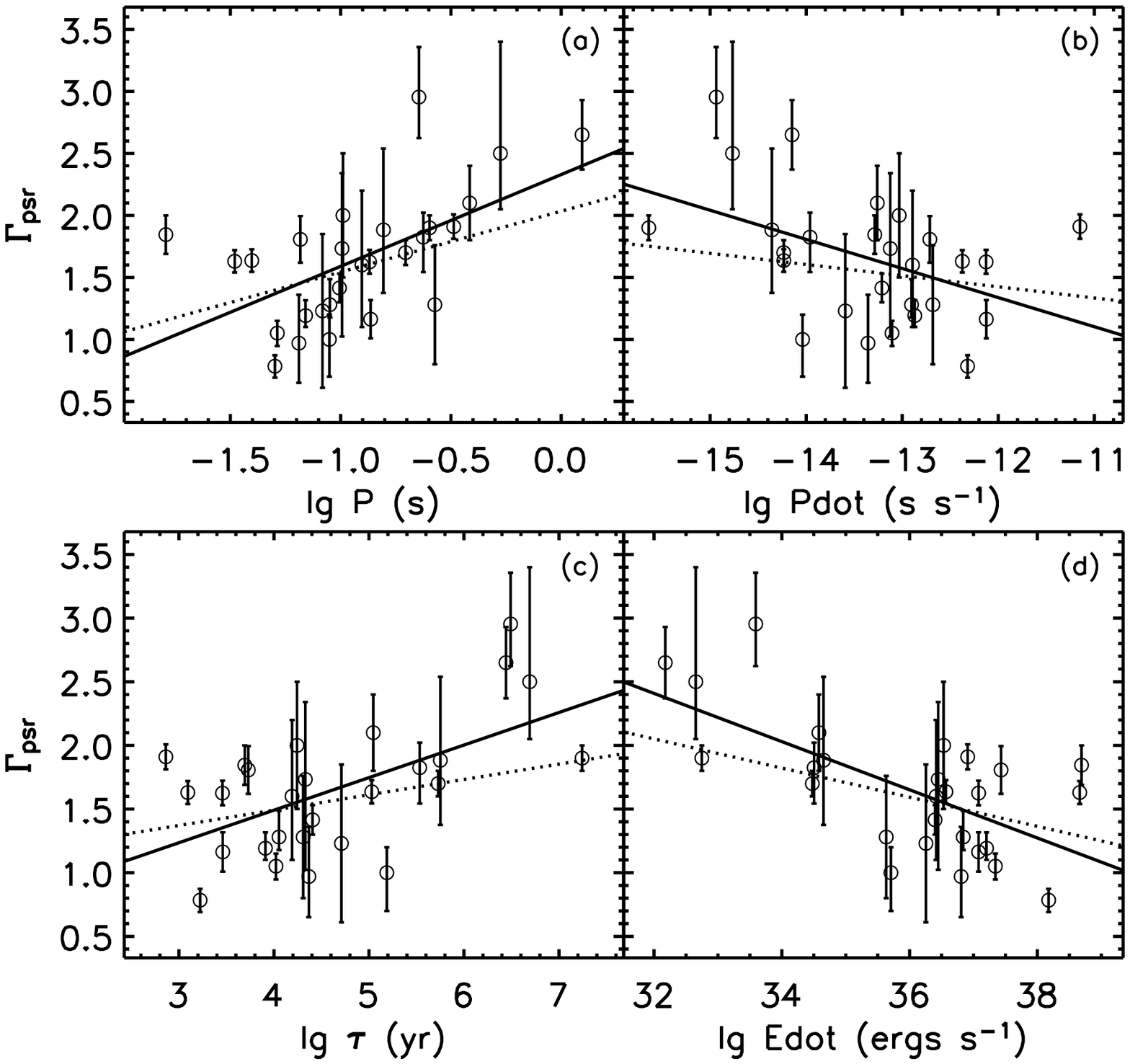}
\caption{Relations of $\Gamma_{\rm psr}$ vs. $P$, $\dot{P}$, $\tau$ and $\dot{E}$.  The
solid lines are the best LSM fit without observational errors considered while the dotted
lines are with errors. The fitting results are also listed in Table 3.}
\label{fig_Gamma_psr}
\end{figure}

We also study the pulsar spectral properties, and check if there is any correlation
between the pulsar spectral index $\Gamma_{\rm psr}$ and the pulsar rotational
parameters. Inspection of Fig. \ref{fig_Gamma_psr}, may indicate the occurrence of a
positive correlation of $\Gamma_{\rm psr}$ with $P$ and $\tau$ and of a negative
correlation of $\Gamma_{\rm psr}$ with $\dot{P}$ and  $\dot{E}$. However, a numerical
test indicates that all these correlations are too weak (the Spearman coefficients
$|r_{\rm s}|$ are all $\lesssim0.60$, see Table 3) for drawing any firm conclusion with
the available data.

\subsection{Correlations between the PWN emission properties
and the pulsar rotational parameters}
\begin{figure}
\includegraphics[scale=1.]{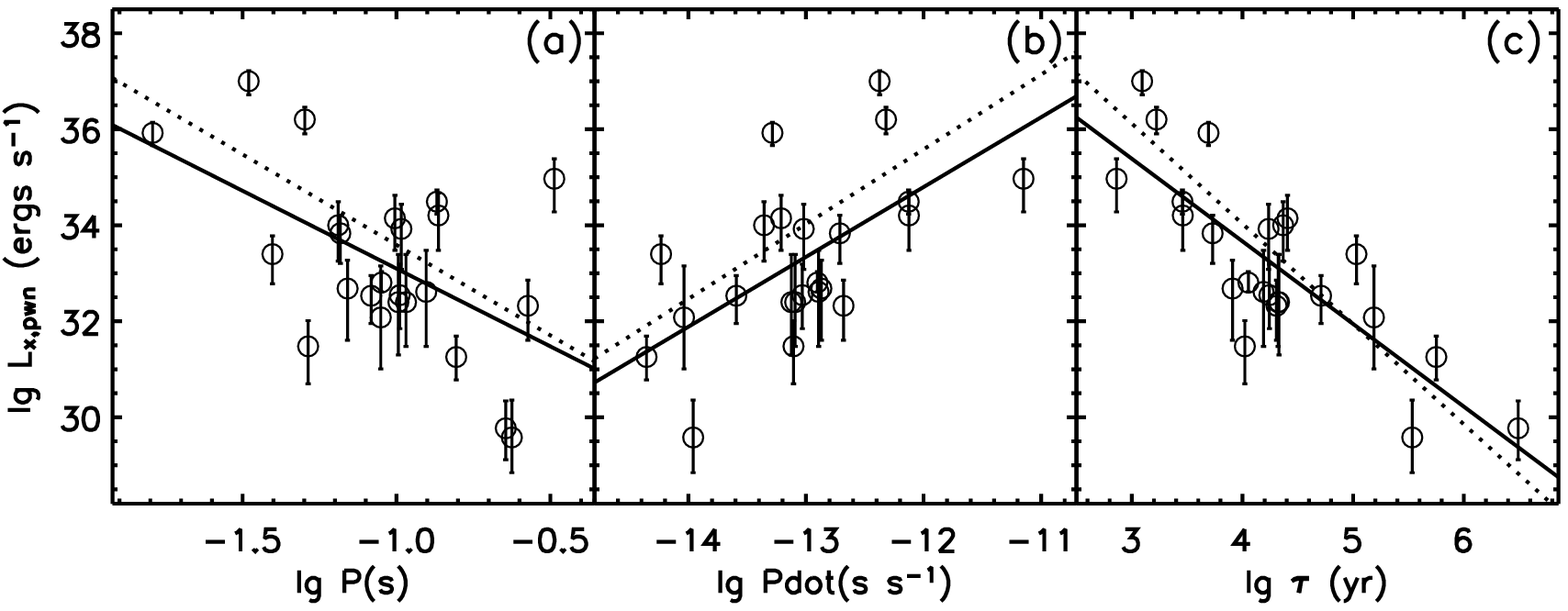}
\caption{Relations of $L_{\rm X, pwn}$ vs. $P$, $\dot{P}$ and $\tau$.  The solid lines are
the best LSM fit without observational errors considered while the dotted lines are with
errors. The fitting results are also listed in Table 3.} \label{fig_Lx_pwn}
\end{figure}

We study here the correlations between the PWN X-ray
luminosity $L_{\rm X,pwn}$ and the pulsar rotational parameters. As shown in Table 3 and
Fig.~\ref{fig_Lx_pwn}, a weak positive correlation between $L_{\rm X,pwn}$ and $\dot{P}$ has
been detected.
Table 3 and Figs~\ref{fig_le} and~\ref{fig_Lx_pwn} also show that $L_{\rm X,pwn}$ is
strongly correlated with $\dot{E}$ and $\tau$.
The linear fits with the observational errors result
\[
  L_{\rm X,pwn} = 10^{-19.6\pm3.0}\dot{E}^{1.45\pm0.08} (\chi^2=2.7)
\]
\[
  L_{\rm X,pwn} = 10^{42.4\pm0.5}\tau^{-2.1\pm0.1} (\chi^2=5.0)
\]
Again, $L_{\rm X,pwn}$ and $\dot{E}$ are in units of erg~s$^{-1}$. Like for the emission
from RPPs, the adopted linear model in the logarithm of the data does not provide a
statistically acceptable fit.

Trying a linear fit which does not account for the uncertainties on the values of $L_{\rm X,pwn}$, we obtain
\[
  L^{*}_{\rm X,pwn} = 10^{-14.9\pm6.0}\dot{E}^{1.3\pm0.2}
\]
\[
  L^{*}_{\rm X,pwn} = 10^{40.5\pm1.1}\tau^{-1.7\pm0.3}
\]
We note that the slope of the relation $L^{*}_{\rm X,pwn}\propto \dot{E}^{1.3}$ is
somewhat different from that of the pulsar, $L^{*}_{\rm X,psr}\propto\dot{E}$. The same
holds true comparing $L_{\rm X,pwn}\propto \dot{E}^{1.45}$ with $L_{\rm
X,psr}\propto\dot{E}^{0.92}$. It is worth noting that, as seen in Fig. \ref{fig_le}, the
scatterings in the relation of $L_{\rm X,psr}$ versus $\dot{E}$ and that of the PWNe are
comparable and both are large.

\begin{figure}
\includegraphics[scale=.7]{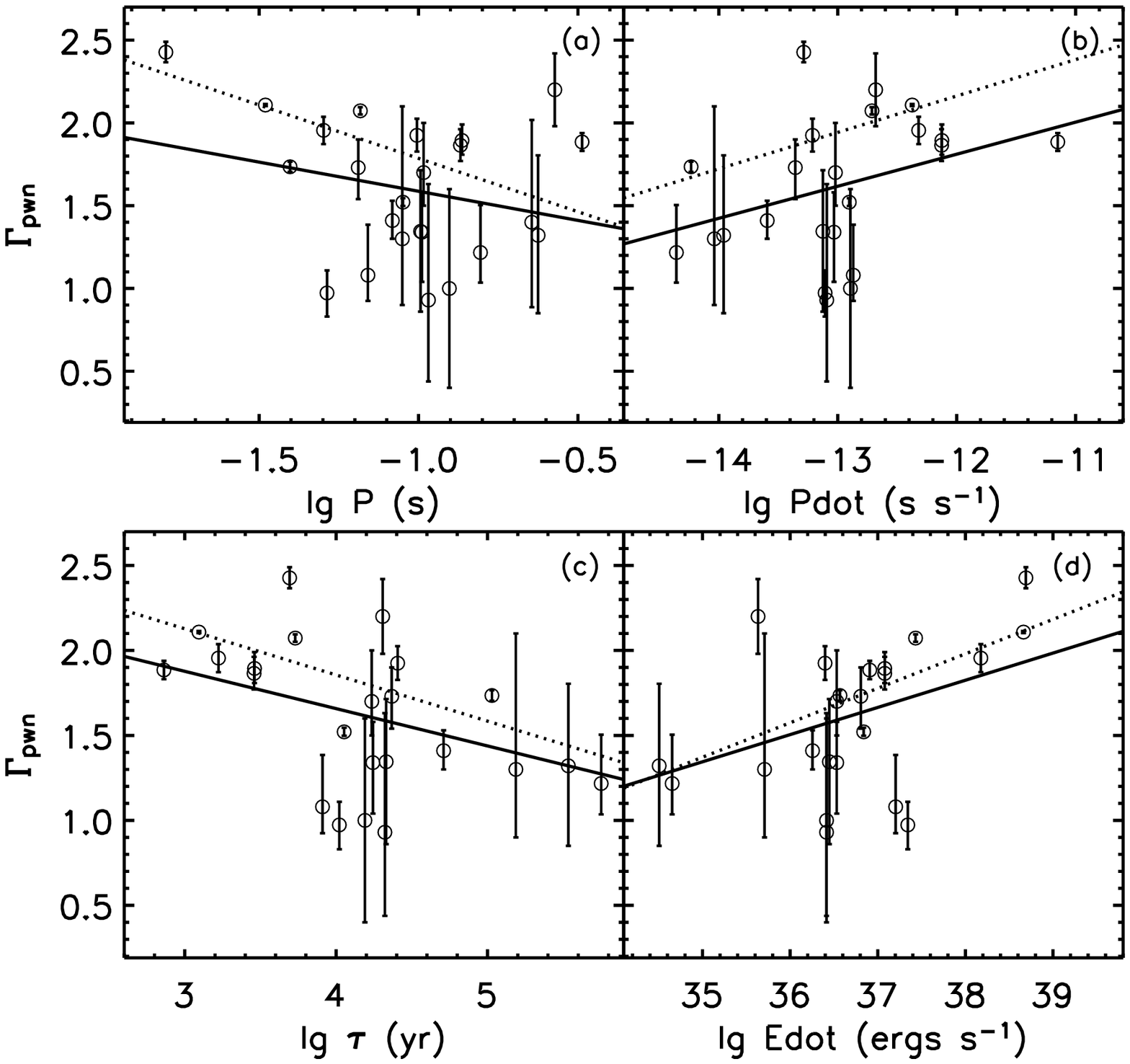}
\caption{Relations of $\Gamma_{\rm pwn}$ vs. $P$, $\dot{P}$, $\tau$ and $\dot{E}$. The
solid lines are the best LSM fit with observational errors considered while the dotted
lines are without errors. The fitting results are also listed in Table 3.}
\label{fig_Gamma_pwn}
\end{figure}

\begin{figure}
\includegraphics[scale=.6]{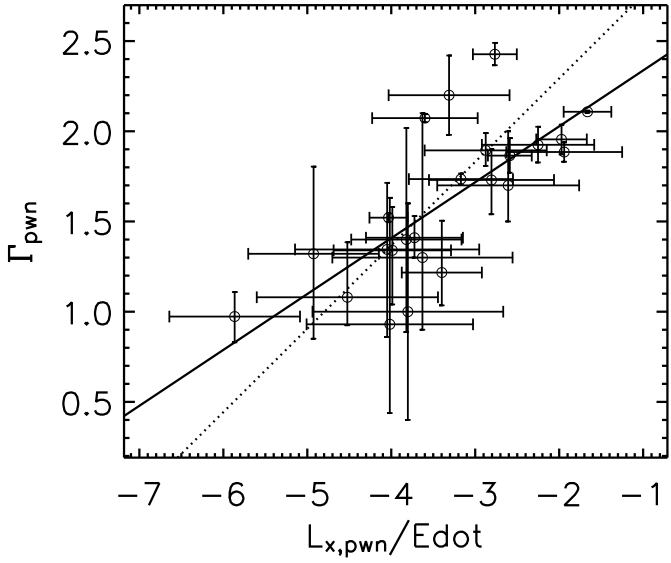}
\caption{The relation between $\Gamma_{\rm pwn}$ and $L_{\rm X,pwn}/\dot{E}$. The solid
line is the best LSM fit without observational errors considered while the dotted line is
with errors. The fitting results are also listed in Table 3.} \label{fig_Gpwn_lx_edot}
\end{figure}

As seen in Fig~\ref{fig_Gamma_pwn} and Table 3, there is no evidence for strong
correlations between $\Gamma_{\rm pwn}$ and the pulsar rotational parameters. The
Spearman rank test also supports this eye-ball study. However in Figs
\ref{fig:Lxvs.Gamma} and \ref{fig_Gpwn_lx_edot} we see obvious positive correlations
between the photon index $\Gamma_{\rm pwn}$ and  the X-ray luminosity $L_{\rm X,pwn}$
or the X-ray conversion efficiency $L_{\rm X,pwn}/\dot{E}$. We will discuss the physical
implication in \S 4.

\subsection{Correlations between non-thermal emission properties of RPPs and PWNe}
\begin{figure}
\includegraphics[scale=.6]{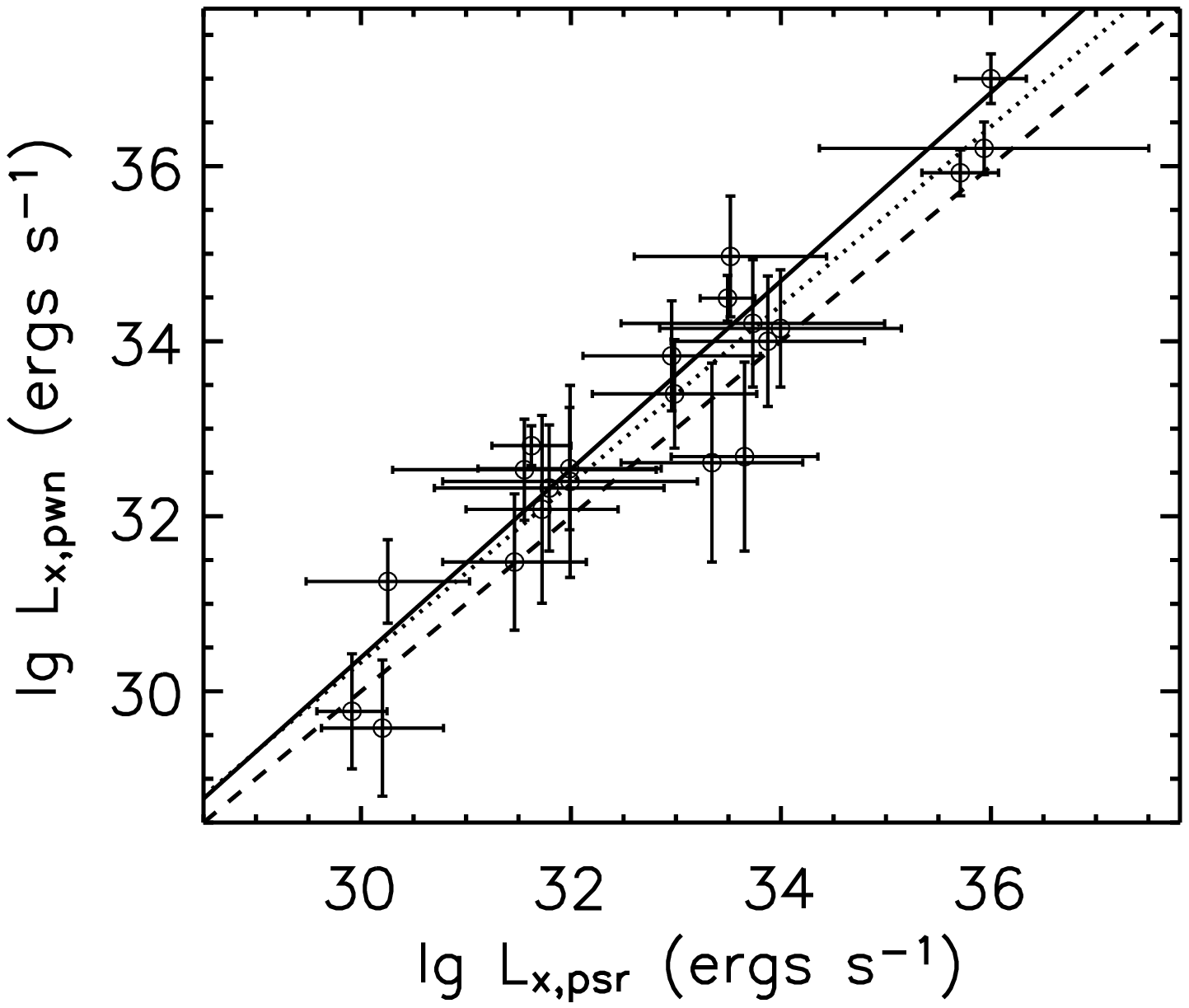}
\includegraphics[scale=.6]{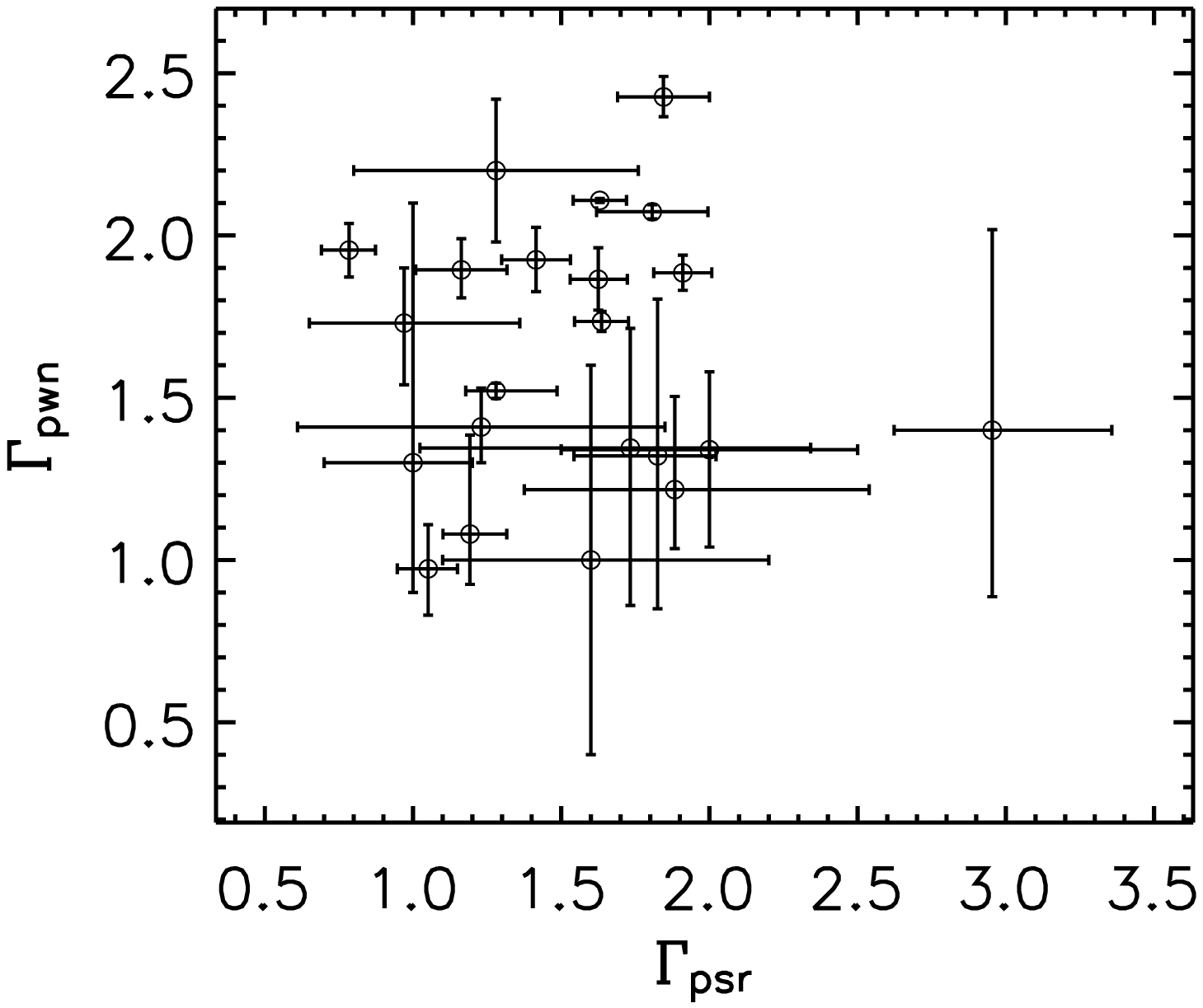}
\caption{Left: the relation of $L_{\rm X, psr}$ vs. $L_{\rm X, pwn}$. The solid lines are
the best LSM fit with the observational errors considered while the dotted lines are
without errors. The fitting results are also listed in Table 3. The dash line marking the
case of $L_{\rm X, psr}=L_{\rm X, pwn}$ is for comparison. Right: the relation of
$\Gamma_{\rm psr}$ vs. $\Gamma_{\rm pwn}$. } \label{fig:pulvsPWN}
\end{figure}

For those samples with both the RPP and the PWN non-thermal X-ray emission measured, we
test the correlations between them. As shown in Fig \ref{fig:pulvsPWN}, a strong
correlation between the X-ray luminosity of RPPs and that of PWNe appears in our samples,
while no correlation shown between their photon indices.

The correlation test between two luminosities gives $r_{\rm s}=0.91$ and $p_{\rm
s}<10^{-4}$, and trying a linear fit to the relation leads to $L_{\rm
X,pwn}=10^{-1.9\pm3.2}L_{\rm X,psr}^{1.1\pm0.1}$. Given the strong positive
correlations of $L_{\rm X,psr}$ and $L_{\rm X,pwn}$ versus $\dot{E}$ separately, a
strong correlation between the two luminosities might be naturally expected. However,
the slope of the relation between the two luminosities is somewhat different from that
expected from the previous two relations, $L_{\rm X,psr}\propto\dot{E}$ and $L_{\rm
X,pwn}\propto\dot{E}^{1.3}$. This may be because that the samples used are different.
For example, those data points with $\dot{E}\la10^{34}\rm ergs\,s^{-1}$ are not
included for the relation of $L_{\rm X,pwn}$ versus $\dot{E}$ since no obvious PWNe
were detected, and they seem to result in a smaller slope for $L_{\rm X,psr}$ versus
$\dot{E}$ relation (Fig \ref{fig_le}). However, it should be noted that both $L_{\rm
X, psr}$ and $L_{\rm X, pwn}$ are actually modulated by the source distance.
The above correlation might be due to the effect of distance modulation.

\begin{figure}
\includegraphics[scale=.6]{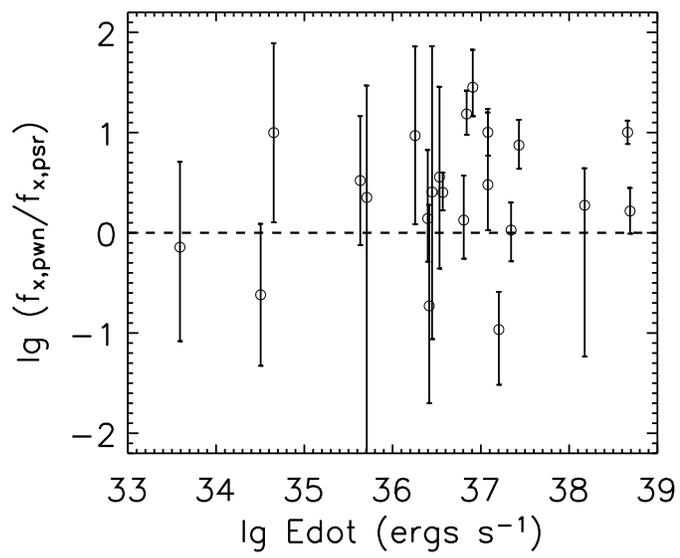}
\caption{ The relation of $f_{\rm X, pwn}/f_{\rm X, psr}$ vs $\dot{E}$. The dash line
indicates $f_{\rm X, pwn} = f_{\rm X, psr}$.} \label{fig:flux_ratio}
\end{figure}

In our samples, the X-ray luminosity ratio between PWNe and RPPs, as shown by
$f_{\rm X,pwn}/f_{\rm X,psr}$ in Fig. \ref{fig:flux_ratio}, varies in the range of
$0.1-30$, about 2 orders of magnitude, and PWNe are generally brighter than their related
RPPs, typically $L_{\rm X,pwn}/L_{\rm X,psr}\sim1-10$. Fig. 9 also tells us that
a more energetic pulsar does not tend to transfer a bigger fraction
of $\dot{E}$ into PWN emission than to the pulsar emission, and vice versa.

Gotthelf (2003) reported a linear relation between $\Gamma_{\rm psr}$ and $\Gamma_{\rm
pwn}$ for the Crab-like pulsars, but we find no correlation between them in our samples
(Fig. \ref{fig:pulvsPWN}). We also check the relation using the samples of Gotthelf
(2003), and a similar relation appears. These might suggest that the linear relation
exists only in the Crab-like pulsars, which are very young.

\section{Discussions}

\subsection{Non-thermal X-ray luminosities of RPPs and PWNe}
The detected X-ray emission from the RPPs and their PWNe is powered by the pulsar
rotation energy. In our sample, the conversion efficiency of $\dot{E}$ to the 2-10 keV
X-ray emission varies in the range of $(L_{\rm X,psr}+L_{\rm
X,pwn})/\dot{E}\sim10^{-6}-10^{-1}$, with the mean at $\sim10^{-3}$, so usually only a
small fraction of the spin-down power goes into the non-thermal X-ray emission. The
non-thermal X-ray emission in a source is usually dominated by the PWN rather than the
pulsar, and on average, $\langle L_{\rm X,pwn}/L_{\rm X,psr}\rangle\sim10$. This implies
that the relations of the pulsar X-ray luminosity and the spin-down power obtained in the
previous works using the low spatial resolution observations at $>2$keV might be
dominated by the PWN emission.

In this work we can separately analyze the luminosity and the spin-down power
relations for the pulsars and their PWNe, thanks to the high spatial resolution of the
observations. A strong positive correlation between
$L_{\rm X,psr}$ and  $\dot{E}$ is obtained in this paper,
similar to the previous results, as shown in Fig \ref{fig_le}. We compare our
results with the previous work in the following. However, one should keep in mind that
there are obvious differences in the analysis processes, i.e., we can separate the RPP
and the PWN emission and do not include the MSPs in the sample, while the previous
work did not separate the RPPs and the PWNe and included MSPs in the analysis.

The relations for the X-ray luminosity of the RPPs in 2-10 keV band are $L_{\rm
X,psr}\propto \dot{E}^{0.92\pm0.04}$ (uncertainties on $L_{\rm X,psr}$ included) and
$L^{*}_{\rm X,psr}\propto \dot{E}^{1.0\pm0.1}$ (not accounting for the uncertainties on
$L_{\rm X,psr}$), respectively. They are roughly in agreement with the scaling found by
Becker \& Tr\"umper (1997), who used the X-ray luminosity in the 0.1-2.4 keV band. The
$L_{\rm X}-\dot{E}$ relations obtained by Saito(1998) and Possenti et al. (2002) appear
steeper than both our derived relations for RPPs, but they are roughly consistent with
both $L_{\rm X,pwn}\propto \dot{E}^{1.45\pm0.08}$ and $L^{*}_{\rm X,pwn}\propto
\dot{E}^{1.3\pm0.2}$ (Fig \ref{fig_le}).  This may suggest that their relations could
also be influenced by the PWN emission due to the lower spatial resolution. The relations
$L_{\rm X,pul} \propto\dot{E}^{1.2\pm0.08}$ and $L_{\rm X,npul} \propto
\dot{E}^{1.4\pm0.1}$, obtained by Cheng et al. (2004) with {\sl ASCA} data, are similar
to our results.

We have shown that the weak negative correlation of $L_{\rm X,psr}$ versus $P$ and the weak
positive correlation of $L_{\rm X,psr}$ versus $\dot{P}$ lead to a strong positive
correlation between $L_{\rm X,psr}$ and $\dot{E}$ in our MSP-excluding RPP samples. On
the other hand in the previous work including the MSPs and the normal RPPs, the
correlations between the X-ray luminosity (might include the PWN emission) and $P$ or
$\dot{P}$ disappear whereas a trend of $L_{\rm X}$ versus $\dot{E}$ is still  there
(e.g., Possenti et al 2002). Moreover, the MSP samples alone also obey a similar
correlation (Possenti et al. 2002). All these factors together strongly suggest that the
X-ray luminosities of RPPs, including the MSPs, are only dependent on their spin-down
powers.

Although there is a strong correlation between $L_{\rm X,psr}$ and $\dot{E}$, the
scattering in this relation is large and the linear fit with the observational errors
included usually gives a statistically unacceptable result, as suggested by Possenti et
al. (2002). $L_{\rm X,psr}$ at given $\dot{E}$ may spread over 2-4 orders of magnitude,
as seen in Fig \ref{fig_le}. The uncertainty in the distance determination and the
momenta of inertia are not expected to lead to such a large span, so other intrinsic
factors may work, e.g., the viewing angle effect, etc. The scattering in the $L_{\rm
X,pwn}-\dot{E}$ relation is comparably large, which is somewhat strange, since the PWN
emission is less influenced by the viewing angles.

\subsection{Non-thermal X-ray spectra of the RPPs}

There are mainly two scenarios to produce the non-thermal X-rays in the magnetospheres of
pulsars. The outer gap scenario (e.g., Cheng et al. 1998; Wang et al. 1998; Cheng \&
Zhang 1999) produces a downward synchrotron-curvature cascade, where the secondary
electrons/positions produce X-rays by synchrotron emission. Another scenario is the polar
gap scenario, e.g., Zhang \& Harding (2000) proposed the ``full polar cap cascade", where
the non-thermal X-rays are produced by resonant inverse Compton (IC) scattering off the
thermal X-ray photons. In both scenarios the $L_{\rm X, psr}\propto\dot{E}$ relation is
generally predicted, although the X-ray spectra are not easy to understand.

We note that a significant fraction of pulsars with very soft spectral indices,
$\Gamma_{\rm psr}\sim2-3$, may pose questions on current models. In a cascade, the
monoenergetic primary electrons emit monoenergetic curvature photons, which subsequently
turn into still monoenergetic pairs in a soft photon bath. The fast energy loss of the
secondary pairs in the magnetic field produces synchrotron emission, which have an photon
spectrum with a power law index $\Gamma_1=1.5$. If the cascade continues the photons
produce next-generation pairs and then synchrotron photons with index
$\Gamma_2=1+\Gamma_1/2=1.75$; furthermore, $\Gamma_3=1+\Gamma_2/2=1.875$... So the
indices will never be bigger than 2, in contrast with the soft spectra. Actually this
discussion could also work if IC rather than synchrotron emission is involved since the
index of synchrotron and IC emission is equal for the same energy distribution of pairs.
As for the polar gap scenarios, the synchrotron emission at X-rays is weak because the
secondary pairs with small pitch angles produce synchrotron emission well above the
cyclotron frequency in the strong pulsar magnetic field, typically $\sim100$~keV. Zhang
\& Harding (2000) proposed that the X-ray emission is dominated by the low energy tail in
the resonant IC emission. However this tail may be hard with index $\Gamma<2$, as shown
in some Monte Carlo simulations (e.g., Fang \& Zhang 2006), although the cases might be
more complicated when more factors such as the viewing angle are taken into account.

Wang \& Zhao (2004) reported the possible negative correlations between
$\Gamma_{\rm psr}$ and $\dot{\Omega}$ and between $\Gamma_{\rm psr}$ and ${\zeta}$,
where $\zeta$ is the generation order parameter characterizing a pulsar under the
scheme of cascade processes (Zhao et al. 1989; Lu et al. 1994; Wei et al. 1997). A
similar negative correlation between $\Gamma_{\rm psr}$ and ${\zeta}$ in anomalous
X-ray pulsars and softer gamma-ray repeaters had also been reported (Marsden \& White
2001; Lu et al. 2003), suggesting that a common mechanism may operate in both normal
and anomalous pulsars. These observational results seem in contrast with the predicted
positive correlation between $\Gamma_{\rm psr}$ and $\zeta$ by Lu et al. (1994). Here,
we also check the relation between $\Gamma_{\rm psr}$ and the generation order
parameter $\zeta_3= 1 + (0.6 - (11/14){\rm log}P + (2/7){\rm log}\dot{P}_{15})/1.3$
(Eq. 6 of Wang \& Zhao 2004) and list the correlation results in Table 3. It turns out
that although there may be some hints of such a negative correlation in our sample,
the correlation tests do not support it strongly, $r_{\rm s}=-0.5$.  So the current
data are not good enough to test the theoretical predictions of Lu et al. (1994).

\subsection{X-ray spectra of PWNe}

In the standard Kennel \& Coroniti (1984ab) model for the Crab nebula, the young Crab
pulsar loses its rotational energy predominantly in the form of a highly relativistic
particle wind, which encounters with the surrounding medium in a termination shock and
become visible by synchrotron emission downstream from the shock. In this context
the energy in the relativistic wind is transferred into post shock magnetic field and
accelerated particles with energy distribution of $N_e(E_e)\propto E_e^{-p}$. Chevalier
(2000) discussed the PWN spectra with emphasis on the cooling of the X-ray emitting
electrons, which leads to a steeper index $p+1$ for high energy and fast-cooling electrons,
and hence a spectral transition of synchrotron photons from $(p+1)/2$ in the slow-cooling
regime to $(p+2)/2$ in the fast-cooling regime at break frequency ($\nu_c$).

The data fit with a single power law model to the indeed broken power law would result in a
spectral index always in the range of $(p+1)/2\leq\Gamma_{\rm pwn}\leq(p+2)/2$. An
electron index value $p\approx2.2$ is generally obtained in theoretical works on particle
acceleration in relativistic collisionless shocks, by both numerical calculations (e.g.,
Achterberg et al. 2001) and analytic analysis (e.g., Keshet \& Waxman 2005), and also
inferred from observation in other kinds of astrophysical relativistic shocks, e.g., GRB
afterglows (e.g., Freedman \& Waxman 2001). Our results show an narrow index range of
$1.5\la\Gamma_{\rm pwn}\la2.1$ unless one source with somewhat higher value $\sim2.5$,
suggesting an electron index of $p\sim2.2$. This consistence with particle wind models
gives a strong support to the Fermi-shock acceleration in PWNe.

We show that the PWN spectral parameters are not strongly correlated with the pulsar
rotational parameters (Fig \ref{fig_Gamma_pwn}). Gotthelf (2003) reported the correlation
between $\Gamma_{\rm psr}$ and $\Gamma_{\rm pwn}$ for nine Crab-like pulsars. Our studies
show that such a correlation is probably not a common property for all RPPs. Therefore,
the electron spectrum and its evolution in a PWN are not determined by the central
pulsar, consistent with wind models where the emission comes from a relativistic shock
between wind and environment interaction.

The relation of $\Gamma_{\rm pwn}$ with  PWN luminosity $L_{\rm X,pwn}$ and the
conversion efficiency $L_{\rm X,pwn}/\dot{E}$ (Figs~\ref{fig:Lxvs.Gamma} and
\ref{fig_Gpwn_lx_edot} and Table 3) could be understood qualitatively in the framework of
pulsar wind models taking into account the electron cooling effect on spectral profile
(e.g., Chevalier 2000). If pulsar loses most of its rotation energy through particle
winds, then higher $\dot{E}$ corresponds to stronger cooling and hence lower spectral
break $\nu_c$, which further means a larger index $\Gamma_{\rm pwn}$ in a fixed
observational energy range. In the meantime, a higher $\dot{E}$ corresponds to a larger
$L_{\rm X,pwn}$, no matter $\nu_c$ is below or above the observational range, and
corresponds to constant X-ray conversion efficiency for fast cooling regime ($\nu_c$
below observed range) or larger $L_{\rm X,pwn}/\dot{E}$ for slow cooling regime ($\nu_c$
above observed range). Therefore we have softer PWN spectra (larger $\Gamma_{\rm pwn}$)
for more luminous PWNe (larger $L_{\rm X,pwn}$) and higher energy conversion efficiency
($L_{\rm X,pwn}/\dot{E}$). This consistence supports the wind-shock model for PWNe. In
this context, the transition of $\Gamma_{\rm pwn}$ from high to low values in Fig.
\ref{fig:Lxvs.Gamma} suggests that the spectral break $\nu_c$ locates at 2-10 keV for
$L_{\rm X,pwn}\sim10^{33}\rm ergs\, s^{-1}$. This may give constraint to wind model
parameters.

\section{Conclusions}

In this work, using the available samples of 27 RPPs and 24 PWNe observed by {\sl
Chandra} and {\sl XMM-Newton}, we obtain the non-thermal X-ray spectral properties, i.e.,
luminosities and spectral indices, of RPPs and PWNe separately. We then analyze their
distribution and correlation with each other and with pulsar rotational parameters.
\begin{itemize}

\item As to the correlations we find: (1) $L_{\rm X,psr}$ and $L_{\rm X,pwn}$
display a strong correlation with both  $\dot{E}$ and $\tau$;  (2) $L_{\rm X,psr}$ also
shows a possible weaker correlation with $P$ and $\dot{P}$, whereas  $L_{\rm X,pwn}$
manifests a similar weak correlation with $\dot{P}$ only; (3) $\Gamma_{\rm pwn}$ is
positively correlated with $L_{\rm X,pwn}$ and the efficiency of conversion of rotational
energy loss in X-ray luminosity $L_{\rm X,pwn}/\dot{E}$.

\item Trying to fit the logarithm of the data with a simple linear fit, we find: $L_{\rm
X,psr}=10^{-0.8\pm1.3}\dot{E}^{0.92\pm0.04}$ and $ L_{\rm X,pwn} =
10^{-19.6\pm3.0}\dot{E}^{1.45\pm0.08}$. However, both the fits are statistically
unacceptable. Not accounting for the uncertainties on the observed luminosity, the
aforementioned relations become $L^{*}_{\rm X,psr}=10^{-4.2\pm3.7}\dot{E}^{1.0\pm0.1} $ and
$L^{*}_{\rm X,pwn} = 10^{-14.9\pm6.0}\dot{E}^{1.3\pm0.2}$, respectively.
Since the scatter in the relation
for PWN (whose emission should be less affected by viewing angle) is comparably larger
than that for RPPs, the scatter in the relation is more probably intrinsic to the
sources.

\item The PWN X-ray luminosity is typically 1 to 10 times larger than that from the
underlying pulsar.

\item The pulsar photon index spans a range of $1\la\Gamma_{\rm
  psr}\la3$. A significant fraction of RPPs with low $\dot{E}$ show soft spectra of $\Gamma_{\rm psr}>2$, which seems not
  consistent with the current models and urges for further investigation of the non-thermal X-ray emission mechanisms of pulsars.

\item The PWN spectral properties are consistent with the particle wind model: the photon
index range $1.5\la\Gamma_{\rm pwn}\la2$ is
  consistent with that expected from the shock-accelerated electrons of index $p\sim2$;
  the correlations of $\Gamma_{\rm pwn}$ with
  $L_{\rm X,pwn}$ and the conversion efficiency $L_{\rm X,pwn}/\dot{E}$ are
  consistent with the wind
  model; no correlation between $\Gamma_{\rm pwn}$ and the pulsar rotational parameters
  also implies that the cooling process is not related to the center pulsars but to the
  interaction of the pulsar wind with its environment.

\end{itemize}

\section*{Acknowledgments}
We thank the referee for very thorough comments. We are grateful to S. N. Zhang, L. M.
Song, J. M. Wang, G. J. Qiao, B. Zhang and H. G. Wang for helpful discussion. We thank
Prof. Wielebinski and Dr. Jessner for critically reading the manuscript and giving many
valuable suggestions. XHL sincerely thanks Prof. Wielebinski for the financial support
during her stay at MPIfR, and thanks J.L. Han for warm hospitality during her stay at
NAOC. This work is supported by the National Science Foundation of China through grants
10573017, 10533020 and 10473010.

\bibliographystyle{aa}

\begin{table}
\caption[]{Rotational parameters and emission properties of 27 pulsars}
\label{table1}
\begin{tabular}{llll@{\hspace{0.15cm}}l@{\hspace{0.2cm}}l@{\hspace{0.15cm}}ll@{\hspace{0.15cm}}ll@{\hspace{0.15cm}}l@{\hspace{0.15cm}}l}
\hline\hline
PSR Name     & $P$    & $\dot{P}$          & $N_{\rm H} $ & $d$        & $\Gamma_{\rm psr}$     & $f_{\rm X}$ (2-10keV)            & $L_{\rm x,psr}$                  & Det. & Ref.    \\
             & $\rm s$ & $\rm s\,s^{-1}$    & $10^{22}\rm cm^{-2}$ & kpc        &                        & $\rm ergs\,s^{-1}\,cm^{-2}$      & $\rm ergs\,s^{-1}$               &      &                     \\
\hline
J0205+6449   & 0.066  & $1.939\times10^{-13}$ & 0.442                  &        3.2$_{   -1.6   }^{+    1.6   }$ & $    1.8_{   -0.2}^{+    0.2}$ & $ 7.4_{    -3.0}^{+     4.7}\times10^{-13}$ & $9.1_{-7.8 }^{+24.3 }\times10^{32}$ & C    & T,  1     \\
B0355+54     & 0.156  & $4.397\times10^{-15}$ & $    0.2_{   -0.1}^{+    0.2}$ &        1.0$_{   -0.2   }^{+    0.2   }$ & $    1.9_{   -0.5}^{+    0.7}$ & $ 1.4_{    -1.0}^{+     3.7}\times10^{-14}$ & $1.8_{-1.5 }^{+7.7  }\times10^{30}$ & C    & T,  2     \\
B0531+21     & 0.033  & $4.228\times10^{-13}$ & 0.345                  &        2.0$_{   -0.5   }^{+    0.5   }$ & $   1.63_{  -0.09}^{+   0.09}$ & $ 2.1_{    -0.4}^{+     0.5}\times10^{-09}$ & $1.0_{-0.5 }^{+0.9  }\times10^{36}$ & X    &   3,  4   \\
J0537$-$6910 & 0.016  & $5.178\times10^{-14}$ & 0.55                   &       50.0$_{  -10.0   }^{+   10.0   }$ & $    1.8_{   -0.2}^{+    0.2}$ & $ 1.7_{    -0.5}^{+     0.8}\times10^{-12}$ & $5.1_{-2.9 }^{+5.5  }\times10^{35}$ & C    & T,  5     \\
B0540$-$69   & 0.050  & $4.791\times10^{-13}$ & $   0.43_{  -0.06}^{+   0.18}$ &       50.0$_{  -10.0   }^{+   10.0   }$ & $   0.78_{  -0.09}^{+   0.09}$ & $ 2.9_{    -1.3}^{+    70.5}\times10^{-12}$ & $8.6_{-5.6 }^{+307.4}\times10^{35}$ & C    & T,  5     \\
\\
B0628$-$28   & 1.244  & $7.123\times10^{-15}$ & $   0.13_{  -0.02}^{+   0.03}$ &        1.4$_{   -0.6   }^{+    0.6   }$ & $    2.6_{   -0.3}^{+    0.3}$ & $ 4.0_{    -1.7}^{+     2.6}\times10^{-15}$ & $1.0_{-0.8 }^{+2.3  }\times10^{30}$ & C    &   6,  7   \\
J0633+1746   & 0.237  & $1.097\times10^{-14}$ & 0.001                  &       0.25$_{  -0.06   }^{+   0.12   }$ & $    1.8_{   -0.3}^{+    0.2}$ & $ 2.1_{    -1.0}^{+     1.6}\times10^{-13}$ & $1.6_{-1.1 }^{+4.5  }\times10^{30}$ & C    & T,  8     \\
B0656+14     & 0.385  & $5.500\times10^{-14}$ & $  0.043_{ -0.002}^{+  0.002}$ &       0.29$_{  -0.03   }^{+   0.03   }$ & $    2.1_{   -0.3}^{+    0.3}$ & $ 9.6_{    -5.5}^{+     7.6}\times10^{-14}$ & $9.5_{-6.2 }^{+11.6 }\times10^{29}$ & X    &   9, 10   \\
B0823+26     & 0.531  & $1.709\times10^{-15}$ & $0.0_{-0.0}^{+0.088}$  &        0.3$_{   -0.1   }^{+    0.1   }$ & $    2.5_{   -0.5}^{+    0.9}$ & $ 2.6_{    -2.0}^{+     4.1}\times10^{-15}$ & $3.6_{-3.3 }^{+14.8 }\times10^{28}$ & X    &  11,  7   \\
B0833$-$45   & 0.089  & $1.250\times10^{-13}$ & 0.017                  &       0.29$_{  -0.05   }^{+   0.08   }$ & $    1.3_{   -0.1}^{+    0.2}$ & $ 4.0_{    -1.5}^{+     2.0}\times10^{-12}$ & $4.2_{-2.4 }^{+5.8  }\times10^{31}$ & C    & T, 12     \\
\\
B0950+08     & 0.253  & $2.298\times10^{-16}$ & $   0.03_{  -0.02}^{+   0.03}$ &      0.262$_{ -0.005   }^{+  0.005   }$ & $    1.9_{   -0.1}^{+    0.1}$ & $ 5.1_{    -1.0}^{+     1.3}\times10^{-14}$ & $4.2_{-0.9 }^{+1.3  }\times10^{29}$ & X    &  11, 13   \\
B1055$-$52   & 0.197  & $5.833\times10^{-15}$ & $  0.027_{ -0.002}^{+  0.002}$ &        0.7$_{   -0.3   }^{+    0.3   }$ & $    1.7_{   -0.1}^{+    0.1}$ & $ 7.8_{    -1.8}^{+     2.8}\times10^{-14}$ & $4.5_{-3.4 }^{+8.0  }\times10^{30}$ & X    &   9, 14   \\
J1124$-$5916 & 0.135  & $7.471\times10^{-13}$ & $   0.28_{  -0.04}^{+   0.04}$ &        6.2$_{   -0.9   }^{+    0.9   }$ & $   1.62_{  -0.10}^{+   0.10}$ & $ 6.8_{    -1.5}^{+     2.0}\times10^{-13}$ & $3.1_{-1.4 }^{+2.2  }\times10^{33}$ & C    & T, 15     \\
J1509$-$5850 & 0.089  & $9.170\times10^{-15}$ & $    0.8_{   -0.2}^{+    0.2}$ &        2.6$_{   -1.0   }^{+    1.0   }$ & $    1.0_{   -0.3}^{+    0.2}$ & $ 6.5_{    -3.3}^{+     7.3}\times10^{-14}$ & $5.3_{-4.3 }^{+16.1 }\times10^{31}$ & C    &  16,  7, 14\\
J1617$-$5055 & 0.069  & $1.351\times10^{-13}$ & $    3.3_{   -0.2}^{+    0.3}$ &        3.3$_{   -1.6   }^{+    1.6   }$ & $   1.19_{  -0.09}^{+   0.12}$ & $ 3.4_{    -1.0}^{+     1.3}\times10^{-12}$ & $4.5_{-3.6 }^{+9.4  }\times10^{33}$ & C    & T, 17     \\
\\
B1706$-$44   & 0.102  & $9.298\times10^{-14}$ & 0.55                   &        2.7$_{   -0.9   }^{+    0.9   }$ & $    2.0_{   -0.5}^{+    0.5}$ & $ 1.1_{    -0.8}^{+     3.0}\times10^{-13}$ & $9.7_{-8.4 }^{+54.6 }\times10^{31}$ & C    &  18, 19   \\
J1747$-$2958 & 0.099  & $6.136\times10^{-14}$ & 2.615                  &        5.0$_{   -2.5   }^{+    2.5   }$ & $    1.4_{   -0.1}^{+    0.1}$ & $ 3.3_{    -2.4}^{+     3.3}\times10^{-12}$ & $9.9_{-9.2 }^{+34.9 }\times10^{33}$ & C    & T, 20     \\
B1757$-$24   & 0.125  & $1.279\times10^{-13}$ & $   3.50_{  -1.10}^{+   1.30}$ &        5.2$_{   -2.1   }^{+    2.1   }$ & $    1.6_{   -0.5}^{+    0.6}$ & $ 6.9_{    -4.4}^{+     9.8}\times10^{-13}$ & $2.2_{-1.9 }^{+8.4  }\times10^{33}$ & C    &  21,  7   \\
J1809$-$1917 & 0.083  & $2.554\times10^{-14}$ & 0.72                   &        3.5$_{   -1.4   }^{+    1.4   }$ & $    1.2_{   -0.6}^{+    0.6}$ & $ 2.5_{    -2.0}^{+    11.4}\times10^{-14}$ & $3.6_{-3.4 }^{+36.3 }\times10^{31}$ & C    &  22,  7   \\
J1811$-$1925 & 0.065  & $4.400\times10^{-14}$ & $    2.2_{   -0.6}^{+    0.8}$ &        5.0$_{   -2.5   }^{+    2.5   }$ & $    1.0_{   -0.3}^{+    0.4}$ & $ 2.5_{    -1.3}^{+     2.1}\times10^{-12}$ & $7.5_{-6.6 }^{+23.4 }\times10^{33}$ & C    &  23, 24   \\
\\
B1823$-$13   & 0.101  & $7.506\times10^{-14}$ & $    1.0_{   -0.6}^{+    0.6}$ &        3.9$_{   -1.6   }^{+    1.6   }$ & $    1.7_{   -0.7}^{+    0.6}$ & $ 5.4_{    -4.5}^{+    33.3}\times10^{-14}$ & $9.8_{-9.2 }^{+130.3}\times10^{31}$ & C    & T,  7     \\
J1846$-$0258 & 0.326  & $7.083\times10^{-12}$ & 3.694                  &        6.0$_{   -3.0   }^{+    3.0   }$ & $   1.91_{  -0.10}^{+   0.10}$ & $ 7.7_{    -3.8}^{+     5.0}\times10^{-13}$ & $3.3_{-2.9 }^{+9.0  }\times10^{33}$ & C    & T, 25     \\
B1853+01     & 0.267  & $2.084\times10^{-13}$ & 5.0                    &        2.6$_{   -1.3   }^{+    1.3   }$ & $    1.3_{   -0.5}^{+    0.5}$ & $ 7.7_{    -5.0}^{+    14.2}\times10^{-14}$ & $6.2_{-5.7 }^{+33.6 }\times10^{31}$ & C    &  26, 27   \\
B1929+10     & 0.227  & $1.157\times10^{-15}$ & $   0.24_{  -0.09}^{+   0.10}$ &      0.361$_{ -0.008   }^{+  0.010   }$ & $    3.0_{   -0.3}^{+    0.4}$ & $ 5.3_{    -2.7}^{+     5.5}\times10^{-14}$ & $8.2_{-4.3 }^{+9.5  }\times10^{29}$ & C    & T,  2     \\
J1930+1852   & 0.137  & $7.506\times10^{-13}$ & 1.6                    &        5.0$_{   -2.5   }^{+    2.5   }$ & $    1.2_{   -0.2}^{+    0.2}$ & $ 1.8_{    -1.4}^{+     2.1}\times10^{-12}$ & $5.4_{-5.1 }^{+21.0 }\times10^{33}$ & C    & T, 28     \\
\\
B1951+32     & 0.040  & $5.845\times10^{-15}$ & 0.299                  &        2.0$_{   -1.0   }^{+    1.0   }$ & $   1.64_{  -0.09}^{+   0.09}$ & $ 2.0_{    -0.6}^{+     0.8}\times10^{-12}$ & $9.7_{-8.1 }^{+20.8 }\times10^{32}$ & C    & T, 29     \\
J2229+6114   & 0.052  & $7.827\times10^{-14}$ & $   0.27_{  -0.07}^{+   0.08}$ &        0.8$_{   -0.4   }^{+    0.4   }$ & $   1.05_{  -0.10}^{+   0.10}$ & $ 3.7_{    -0.7}^{+     1.3}\times10^{-13}$ & $2.9_{-2.3 }^{+5.9  }\times10^{31}$ & C    & T, 30     \\
\hline\hline
\end{tabular}
\tablecomments{C: {\sl Chandra}, X: {\sl XMM-Newton}\\The error bars of $f_{\rm X}$ are derived from the errors in the spectral index and the normalization of spectrum fitting, the error bars of $L_{\rm X,psr}$ are derived from the errors of both the fluxes and the distances. If more than one papers is listed in Ref. column, the first one is where the spectral properties are from and the other ones are those where the distance is from. }
 \tablerefs{
T: This paper; [1] Roberts et al.(1993); [2] Chatterjee et al.(2004); [3] Willingale et
al.(2001); [4] Trimble \& Woltjer(1971); [5] Bradley(2007); [6] {\" O}gelman \&
Tepedelenlio{\v g}lu(2004); [7] Cordes \& Lazio(2002); [8] Faherty et al.(2007); [9] De
Luca et al.(2005); [10] Brisken et al.(2003); [11] Becker et al.(2004); [12] Caraveo et
al.(2001); [13] Brisken et al.(2002); [14] Kramer et al.(2003); [15] Gaensler \&
Wallace(2003); [16] Hui \& Becker(2007); [17] Paron et al.(2006); [18] Gotthelf et
al.(2002); [19] Koribalski et al.(1995); [20] Gaensler et al.(2004); [21] Kaspi et
al.(2001); [22] Kargaltsev \& Pavlov(2007); [23] Roberts et al.(2003); [24] Green et
al.(1988); [25] Leahy \& Tian(2007); [26] Petre et al.(2002); [27] Cox et al. (1999);
[28] Lu et al.(2002); [29] Strom \& Stappers(2000); [30] Kothes et al.(2006). }
\end{table}

\begin{table}
\caption[]{Emission properties of 24 PWNe}
\label{table2}
\begin{tabular}{llllllrl}
\hline\hline
PSR Name     & $N_{\rm H} $ & $d$        & $\Gamma_{\rm pwn}$     & $f_{\rm X}$ (2-10keV)            & $L_{\rm X,pwn}$                  & Det. & Ref.                \\
             & $10^{22} \rm cm^{-2}$  & kpc        &                        & $\rm ergs\,s^{-1}\,cm^{-2}$      & $\rm ergs\,s^{-1}$               &      &                     \\
\hline
J0205+6449   & $   0.44_{  -0.01}^{+   0.01}$ &        3.2$_{   -1.6   }^{+    1.6   }$&$   2.07_{  -0.02}^{+   0.02}$ & $ 5.6_{    -0.3}^{+     0.3}\times10^{-12}$ & $6.8_{-5.2 }^{+9.4  }\times10^{33}$ & C    & T         \\
B0355+54     & $    0.2_{   -0.1}^{+    0.2}$ &        1.0$_{   -0.2   }^{+    0.2   }$&$    1.2_{   -0.2}^{+    0.3}$ & $ 1.4_{    -0.7}^{+     1.2}\times10^{-13}$ & $1.8_{-1.2 }^{+3.1  }\times10^{31}$ & C    & T         \\
B0531+21     & $0.345$                &        2.0$_{   -0.5   }^{+    0.5   }$&$  2.108_{ -0.006}^{+  0.006}$ & $ 2.1_{    -0.1}^{+     0.1}\times10^{-08}$ & $1.0_{-0.5 }^{+0.7  }\times10^{37}$ & X    &   1       \\
J0537$-$6910 & $   0.55_{  -0.03}^{+   0.03}$ &       50.0$_{  -10.0   }^{+   10.0   }$&$   2.43_{  -0.06}^{+   0.06}$ & $ 2.8_{    -0.4}^{+     0.4}\times10^{-12}$ & $8.4_{-3.8 }^{+5.6  }\times10^{35}$ & C    & T         \\
B0540$-$69   & $0.43$                 &       50.0$_{  -10.0   }^{+   10.0   }$&$   1.96_{  -0.08}^{+   0.08}$ & $ 5.4_{    -1.1}^{+     1.4}\times10^{-12}$ & $1.6_{-0.8 }^{+1.3  }\times10^{36}$ & C    & T         \\
\\
J0633+1746   & $0.001$                &       0.25$_{  -0.06   }^{+   0.12   }$&$    1.3_{   -0.5}^{+    0.5}$ & $ 5.1_{    -3.3}^{+     9.0}\times10^{-14}$ & $3.8_{-3.1 }^{+19.2 }\times10^{29}$ & C    & T         \\
B0833$-$45   & $0.017$                &       0.29$_{  -0.05   }^{+   0.08   }$&$   1.52_{  -0.02}^{+   0.02}$ & $ 6.2_{    -0.4}^{+     0.4}\times10^{-11}$ & $6.4_{-2.3 }^{+4.4  }\times10^{32}$ & C    & T         \\
J1016$-$5857$\star$ & $    0.3_{   -0.3}^{+    0.7}$ &        3.0$_{   -0.6   }^{+    0.6   }$&$    0.9_{   -0.5}^{+    0.7}$ & $ 2.3_{    -1.9}^{+    13.4}\times10^{-13}$ & $2.5_{-2.2 }^{+21.8 }\times10^{32}$ & C    & T,  2     \\
J1124$-$5916 & $0.281$                &        6.2$_{   -0.9   }^{+    0.9   }$&$   1.86_{  -0.10}^{+   0.10}$ & $ 6.8_{    -1.7}^{+     2.2}\times10^{-12}$ & $3.1_{-1.4 }^{+2.3  }\times10^{34}$ & C    & T         \\
J1509$-$5850 & $    0.8_{   -0.4}^{+    0.9}$ &        2.6$_{   -1.0   }^{+    1.0   }$&$    1.3_{   -0.4}^{+    0.8}$ & $ 1.5_{    -1.5}^{+     8.0}\times10^{-13}$ & $1.2_{-1.2 }^{+13.4 }\times10^{32}$ & C    &   3       \\
\\
J1617$-$5055 & $    3.3_{   -0.2}^{+    0.3}$ &        3.3$_{   -1.6   }^{+    1.6   }$&$    1.1_{   -0.2}^{+    0.3}$ & $ 3.7_{    -2.3}^{+     2.7}\times10^{-13}$ & $4.8_{-4.4 }^{+13.8 }\times10^{32}$ & C    & T         \\
B1706$-$44   & $    0.6_{   -0.2}^{+    0.2}$ &        2.7$_{   -0.9   }^{+    0.9   }$&$    1.3_{   -0.3}^{+    0.2}$ & $ 4.0_{    -2.2}^{+     5.7}\times10^{-13}$ & $3.5_{-2.8 }^{+11.6 }\times10^{32}$ & C    &   4       \\
J1747$-$2958 & $    2.6_{   -0.1}^{+    0.1}$ &        5.0$_{   -2.5   }^{+    2.5   }$&$   1.93_{  -0.10}^{+   0.10}$ & $ 4.6_{    -1.2}^{+     1.6}\times10^{-12}$ & $1.4_{-1.1 }^{+2.8  }\times10^{34}$ & C    & T         \\
B1757$-$24   & $3.5$                  &        5.2$_{   -2.1   }^{+    2.1   }$&$    1.0_{   -0.6}^{+    0.6}$ & $ 1.3_{    -0.9}^{+     3.5}\times10^{-13}$ & $4.1_{-3.8 }^{+26.4 }\times10^{32}$ & C    &   5       \\
J1809$-$1917 & $0.72$                 &        3.5$_{   -1.4   }^{+    1.4   }$&$    1.4_{   -0.1}^{+    0.1}$ & $ 2.3_{    -0.6}^{+     0.8}\times10^{-13}$ & $3.4_{-2.5 }^{+5.6  }\times10^{32}$ & C    &   6       \\
\\
J1811$-$1925 & $    2.1_{   -0.1}^{+    0.1}$ &        5.0$_{   -2.5   }^{+    2.5   }$&$    1.7_{   -0.2}^{+    0.2}$ & $ 3.4_{    -0.8}^{+     1.3}\times10^{-12}$ & $1.0_{-0.8 }^{+2.1  }\times10^{34}$ & C    &   7       \\
B1823$-$13   & $    1.0_{   -0.6}^{+    0.6}$ &        3.9$_{   -1.6   }^{+    1.6   }$&$    1.3_{   -0.5}^{+    0.4}$ & $ 1.4_{    -1.0}^{+     5.3}\times10^{-13}$ & $2.5_{-2.3 }^{+21.7 }\times10^{32}$ & C    & T         \\
J1846$-$0258 & $    3.7_{   -0.1}^{+    0.1}$ &        6.0$_{   -3.0   }^{+    3.0   }$&$   1.89_{  -0.05}^{+   0.05}$ & $ 2.2_{    -0.3}^{+     0.4}\times10^{-11}$ & $9.3_{-7.4 }^{+15.5 }\times10^{34}$ & C    & T         \\
B1853+01     & $5.0$                  &        2.6$_{   -1.3   }^{+    1.3   }$&$    2.2_{   -0.2}^{+    0.2}$ & $ 2.6_{    -0.9}^{+     1.4}\times10^{-13}$ & $2.1_{-1.7 }^{+5.1  }\times10^{32}$ & C    &   8       \\
B1929+10     & $   0.24_{  -0.09}^{+   0.10}$ &      0.361$_{ -0.008   }^{+  0.010   }$&$    1.4_{   -0.5}^{+    0.6}$ & $ 3.8_{    -2.9}^{+     9.6}\times10^{-14}$ & $5.9_{-4.6 }^{+16.1 }\times10^{29}$ & C    & T         \\
\\
J1930+1852   & $   1.64_{  -0.09}^{+   0.10}$ &        5.0$_{   -2.5   }^{+    2.5   }$&$   1.89_{  -0.09}^{+   0.10}$ & $ 5.4_{    -1.3}^{+     1.6}\times10^{-12}$ & $1.6_{-1.3 }^{+3.1  }\times10^{34}$ & C    & T         \\
B1951+32     & $   0.30_{  -0.01}^{+   0.01}$ &        2.0$_{   -1.0   }^{+    1.0   }$&$   1.74_{  -0.03}^{+   0.03}$ & $ 5.1_{    -0.4}^{+     0.4}\times10^{-12}$ & $2.5_{-1.9 }^{+3.5  }\times10^{33}$ & C    & T         \\
J2021+3651$\star$ & $    0.8_{   -0.1}^{+    0.2}$ &        8.0$_{   -4.0   }^{+    4.0   }$&$    1.7_{   -0.2}^{+    0.3}$ & $ 1.1_{    -0.4}^{+     0.5}\times10^{-12}$ & $8.4_{-7.2 }^{+19.2 }\times10^{33}$ & C    &   9       \\
J2229+6114   & $   0.27_{  -0.07}^{+   0.08}$ &        0.8$_{   -0.4   }^{+    0.4   }$&$    1.0_{   -0.1}^{+    0.1}$ & $ 4.0_{    -1.3}^{+     2.0}\times10^{-13}$ & $3.0_{-2.5 }^{+7.3  }\times10^{31}$ & C    & T         \\
\hline\hline
\end{tabular}
\tablecomments{C: {\sl Chandra}, X: {\sl XMM-Newton}\\
$\star$ PSRs J1016$-$5857 and J2021+3651 are not listed in Table 1. Their periods are $P=0.107$ and 0.104~s, respectively, and their period derivatives are $\dot{P}=8.08\times10^{-14}$ and $9.56\times10^{-14} \rm s\,s^{-1}$, respectively.
\\ The error bars of $f_{\rm X}$ are derived from the errors in the spectral index and the normalization of spectrum fitting. The error bars of $L_{\rm X,pwn}$ are derived from the errors of both the fluxes and the distances.}
\tablerefs{ T: This paper; [1] Willingale et al.(2001); [2]Ruiz \& May(1986); [3] Hui \&
Becker(2007); [4] Gotthelf et al.(2002); [5] Kaspi et al.(2001); [6] Kargaltsev \&
Pavlov(2007); [7] Roberts et al.(2003); [8] Petre et al.(2002); [9] Hessels et al.(2004).
}
\end{table}

\begin{table}
\caption[]{Correlation coefficients between parameters}
\label{table3}
\begin{tabular}{rrrrlr@{$\pm$}lr@{$\pm$}lr@{$\pm$}lr@{$\pm$}l}
\hline\hline
$y$         &$x$         &$r$         &$r_{\rm s}$ &$p_{\rm s}$&\multicolumn{2}{c}{$a_{\rm e}$}&\multicolumn{2}{c}{$b_{\rm e}$}&\multicolumn{2}{c}{$a_{\rm ne}$}&\multicolumn{2}{c}{$b_{\rm ne}$}\\\hline
$L_{\rm X,psr}$ & $P$        & $-$0.73    & $-$0.66    & 0.0002     & 27.1       & 0.2        & $-$5.2     & 0.2        & 28.9       & 0.7        & $-$3.6     & 0.7       \\
     & $\dot{P}$  & 0.70       & 0.69       & 0.0001     & 49.7       & 0.8        & 1.29       & 0.06       & 51.0       & 3.9        & 1.4        & 0.3       \\
     & $B$        & 0.41       & 0.46       & 0.0165     & 1.6        & 1.6        & 2.5        & 0.1        & 10.9       & 9.4        & 1.7        & 0.8       \\
     & $\tau$     & $-$0.84    & $-$0.81    & $<10^{-4}$ & 38.1       & 0.3        & $-$1.19    & 0.05       & 38.9       & 0.9        & $-$1.4     & 0.2       \\
     & $\dot{E}$  & 0.89       & 0.82       & $<10^{-4}$ & $-$0.8     & 1.3        & 0.92       & 0.04       & $-$4.2     & 3.7        & 1.0        & 0.1       \\
\hline
$\Gamma_{\rm psr}$ & $P$        & 0.57       & 0.60       & 0.0010     & 2.03       & 0.08       & 0.49       & 0.08       & 2.3        & 0.2        & 0.7        & 0.2       \\
     & $\dot{P}$  & $-$0.45    & $-$0.42    & 0.0274     & 0.3        & 0.3        & $-$0.09    & 0.02       & $-$1.5     & 1.3        & $-$0.23    & 0.09      \\
     & $B$        & $-$0.22    & $-$0.22    & 0.2598     & 2.5        & 0.6        & $-$0.08    & 0.05       & 4.6        & 2.6        & $-$0.2     & 0.2       \\
     & $\tau$     & 0.58       & 0.44       & 0.0205     & 1.0        & 0.1        & 0.12       & 0.02       & 0.5        & 0.3        & 0.26       & 0.07      \\
     & $\dot{E}$  & $-$0.64    & $-$0.51    & 0.0067     & 5.7        & 0.6        & $-$0.11    & 0.02       & 8.5        & 1.6        & $-$0.19    & 0.04      \\
     & $\zeta$    & $-$0.64    & $-$0.50    & 0.0081     & 2.9        & 0.2        & $-$0.54    & 0.08       & 3.8        & 0.5        & $-$0.9     & 0.2       \\
     & $L_{\rm X,psr}$ & $-$0.55    & $-$0.54    & 0.0033     & 5.4        & 0.5        & $-$0.12    & 0.02       & 6.2        & 1.4        & $-$0.14    & 0.04      \\
\hline
\hline
$L_{\rm X,pwn}$ & $P$        & $-$0.54    & $-$0.40    & 0.0519     & 29.8       & 0.3        & $-$3.8     & 0.3        & 29.9       & 1.2        & $-$3.2     & 1.1       \\
     & $\dot{P}$  & 0.66       & 0.61       & 0.0015     & 54.3       & 1.7        & 1.6        & 0.1        & 52.3       & 4.7        & 1.5        & 0.4       \\
     & $B$        & 0.43       & 0.52       & 0.0090     & 21.2       & 3.0        & 1.0        & 0.2        & 11.5       & 9.8        & 1.7        & 0.8       \\
     & $\tau$     & $-$0.82    & $-$0.76    & $<10^{-4}$ & 42.4       & 0.5        & $-$2.1     & 0.1        & 40.5       & 1.1        & $-$1.7     & 0.3       \\
     & $\dot{E}$  & 0.86       & 0.75       & $<10^{-4}$ & $-$19.6    & 3.0        & 1.45       & 0.08       & $-$14.9    & 6.0        & 1.3        & 0.2       \\
     & $L_{\rm X,psr}$ & 0.94       & 0.91       & $<10^{-4}$ & $-$1.9     & 3.2        & 1.08       & 0.10       & $-$0.3     & 2.6        & 1.02       & 0.08      \\
\hline
$\Gamma_{\rm pwn}$ & $P$        & $-$0.25    & $-$0.21    & 0.3216     & 1.14       & 0.05       & $-$0.65    & 0.03       & 1.2        & 0.3        & $-$0.4     & 0.3       \\
     & $\dot{P}$  & 0.38       & 0.40       & 0.0551     & 4.8        & 0.2        & 0.22       & 0.01       & 4.1        & 1.3        & 0.2        & 0.1       \\
     & $B$        & 0.27       & 0.35       & 0.0889     & $-$0.2     & 0.3        & 0.18       & 0.03       & $-$1.5     & 2.4        & 0.2        & 0.2       \\
     & $\tau$     & $-$0.45    & $-$0.45    & 0.0279     & 2.94       & 0.04       & $-$0.27    & 0.01       & 2.5        & 0.4        & $-$0.22    & 0.09      \\
     & $\dot{E}$  & 0.45       & 0.43       & 0.0367     & $-$5.7     & 0.3        & 0.203      & 0.008      & $-$4.3     & 2.4        & 0.16       & 0.07      \\
     & $\Gamma_{\rm psr}$ & $-$0.01    & 0.04       & 0.8554     & 0.2        & 0.3        & 1.1        & 0.2        & 1.6        & 0.3        & $-$0.01    & 0.20      \\
     & $L_{\rm X,pwn}$ & 0.70       & 0.70       & 0.0001     & $-$4.9     & 0.7        & 0.20       & 0.02       & $-$3.8     & 1.2        & 0.16       & 0.04      \\
     & $L_{\rm X,pwn}/\dot{E}$ & 0.73       & 0.75       & $<10^{-4}$ & 3.2        & 0.2        & 0.46       & 0.08       & 2.6        & 0.2        & 0.31       & 0.06      \\
\hline
$f_{\rm X,pwn}/f_{\rm X,psr}$ & $\dot{E}$  & 0.13       & 0.15       & 0.4917     & $-$2.3     & 5.7        & 0.07       & 0.15       & $-$20.6    & 43.4       & 0.7        & 1.2       \\
\hline\hline
\end{tabular}
\tablecomments{$r$ is Pearson correlation coefficient, while $r_{\rm s}$ and $p_{\rm s}$ are Spearman rank correlation coefficient and significance level, respectively. Coefficients $a_{\rm e}$ and $b_{\rm e}$ ($a_{\rm ne}$ and $b_{\rm ne}$) are obtained in the linear fitting with (without) observational errors. }
\end{table}

\end{document}